\documentclass[twocolumn,aps,pra,longbibliography,superscriptaddress,nofootinbib,10pt]{revtex4-1}
\usepackage[latin1,utf8]{inputenc}
\usepackage[english]{babel}
\usepackage{graphicx,dcolumn,bm}
\usepackage{subfigure}
\usepackage{mathrsfs} 
\usepackage{braket}
\usepackage{multirow}
\usepackage{array}
\usepackage{arydshln}
\usepackage{color}
\usepackage[colorlinks,linkcolor=blue,citecolor=blue,hyperindex]{hyperref}
\usepackage{amsfonts}
\usepackage{amssymb}
\usepackage{amsmath}
\usepackage{latexsym}
\usepackage{simplewick}
\usepackage{epstopdf}
\usepackage{multirow}
\usepackage{float}
\usepackage[table]{xcolor}
\usepackage{multibib}
\usepackage{mathrsfs}
\usepackage[sans]{dsfont}

\usepackage[normalem]{ulem}

\usepackage{orcidlink}

\usepackage[mathscr]{eucal}

\usepackage[normalem]{ulem}
\usepackage{epsfig}
\definecolor{Blue}{rgb}{0.0,0.0,1}
\definecolor{Red}{rgb}{1,0.0,0.0}
\definecolor{Green}{rgb}{0,0.5,0.0}
\usepackage{tikz}
\usetikzlibrary{decorations.pathmorphing}
\usetikzlibrary{arrows}
\usetikzlibrary{intersections,shapes.arrows}
\usetikzlibrary{calc}
\usetikzlibrary{quotes,angles}
\usepackage{nicefrac}
\usepackage{pgfplots}
\usepgfplotslibrary{fillbetween}
\pgfplotsset{compat=1.13,colormap={violetnew}{rgb=(0.293416, 0.0574044, 0.529412) rgb=(0.394818,0.233715,0.671945) rgb =(0.49622,0.410025,0.814477) rgb=(0.588672,0.567494,0.910066) rgb=(0.663226,0.687282,0.911765) rgb=(0.73778,0.807069,0.913465) rgb=(0.807267,0.861883,0.894034) rgb=(0.874222,0.884211,0.864039) rgb=(0.941176, 0.906538, 0.834043)}}
\usepgfplotslibrary{groupplots} 
\usepgfplotslibrary[groupplots] 
\usetikzlibrary{pgfplots.groupplots} 
\usetikzlibrary[pgfplots.groupplots] 
\usepgfplotslibrary{statistics}
\usepackage{pgfplotstable}
\tikzset{jumpdot/.style={mark=*,solid},excl/.append style={jumpdot,fill=white},incl/.append style={jumpdot,fill=black}}
\begin{document}

\title{Exponentially accelerated relaxation and quantum Mpemba effect in open quantum systems}
\author{Emerson Lima Caldas\,\orcidlink{0009-0003-5506-6170}}
\affiliation{Programa de P\'{o}s-Gradua\c{c}\~{a}o em F\'{i}sica, Universidade Federal do Maranh\~{a}o, Campus Universit\'{a}rio do Bacanga, 65080-805, S\~{a}o Lu\'{i}s, Maranh\~{a}o, Brazil}
\author{Diego Paiva Pires\,\orcidlink{0000-0003-4936-1969}}
\affiliation{Programa de P\'{o}s-Gradua\c{c}\~{a}o em F\'{i}sica, Universidade Federal do Maranh\~{a}o, Campus Universit\'{a}rio do Bacanga, 65080-805, S\~{a}o Lu\'{i}s, Maranh\~{a}o, Brazil}
\affiliation{Coordena\c{c}\~{a}o do Curso de F\'{i}sica -- Bacharelado, Universidade Federal do Maranh\~{a}o, Campus Universit\'{a}rio do Bacanga, 65080-805, S\~{a}o Lu\'{i}s, Maranh\~{a}o, Brazil}

\begin{abstract}
We investigate the quantum Mpemba effect in the relaxation of open quantum systems whose effective dynamics is described by Davies maps. We introduce a class of unitary transformations constructed from permutation matrices that, when applied to an initial state, (i) suppress the slowest decaying modes of the nonunitary dynamics and (ii) maximize the state's distinguishability from the steady state. The first condition ensures exponentially faster convergence to equilibrium, while the second implies that a quantum system initially further from equilibrium can approach it more rapidly than one that starts closer. This protocol thus realizes a genuine Mpemba effect, and its simulation requires low computational effort. We prove that, for any initial state, there exists a permutation matrix that maximizes its distance from equilibrium with respect to a chosen information-theoretic distinguishability measure. We illustrate our findings for a two-level system, as well as for the nonunitary dynamics of the transverse-field Ising chain and the XXZ chain, each weakly coupled to a bosonic thermal bath. In these cases, the quantum Mpemba effect is demonstrated using the Hilbert-Schmidt distance, quantum relative entropy, and trace distance. Overall, our results provide a versatile framework for engineering a genuine quantum Mpemba effect in Markovian open quantum systems.
\end{abstract}

\maketitle


\section{Introduction}
\label{sec:00000000001}

Intuition suggests that the cooling of a physical system is a monotonic process: The closer the system is to equilibrium, the faster it should reach that state. However, observations dating back to Aristotle over 2000 years ago~\cite{Aristoteles}, along with modern stu\-dies developed by Mpemba and Osborne~\cite{Mpemba_1969} and Kell~\cite{Kell_1969}, reveal counterintuitive behavior: Under certain conditions, initially hotter systems can cool more rapidly than initially colder systems. This phenomenon is known as the Mpemba effect. Since then, it has been observed in various physical systems, including crystalline polymers, manganites, and colloidal suspensions~\cite{doi:10.1021/acs.cgd.8b01250,chaddah2010,kumar2020exponentially}. Despite its widespread occurrence, the underlying mechanism remains the subject of intense debate~\cite{Burridge2016,Bechhoefer2021}. Recent advances in stochastic thermodynamics suggest that initially hotter systems can exploit dynamical shortcuts, accessing regions of state space that allow them to relax more rapidly than colder systems~\cite{Oren-Raz,PhysRevX.9.021060}. We also mention the interplay between Mpemba and Kovacs effects in the so-called time-delayed cooling law~\cite{PhysRevE.109.044149}.
\begin{figure}[!t]
\begin{center}
\includegraphics[scale=0.7]{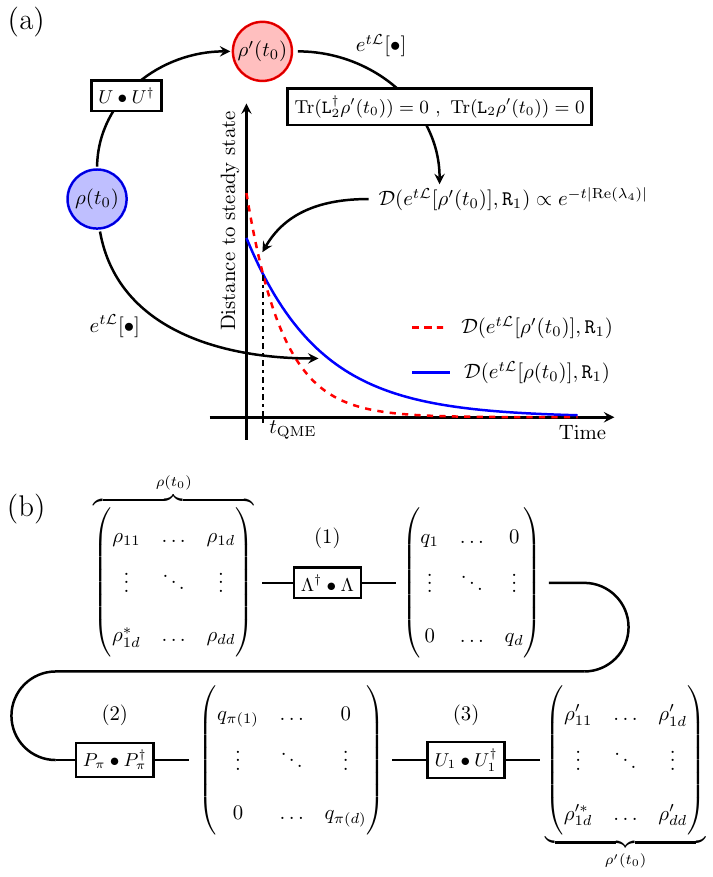}
\caption{(Color online) Overview of the quantum Mpemba effect. (a) We consider a probe state $\rho({t_0})$ that is transformed into ${\rho'}({t_0}) = U\rho({t_0}){U^{\dagger}}$ by means of the unitary matrix $U = {U_1}{P_{\pi}}{\Lambda^{\dagger}}$. This state contributes to eliminate the slowest decaying mode in Davies maps with generator $\mathcal{L}[\bullet]$. Therefore, the relaxation to equilibrium is accelerated exponentially, controlled by the real part of eigenvalue $\lambda_4$. The unitary $U$ maximizes the distance between ${\rho'}({t_0})$ and the steady state ${\texttt{R}_1}$ such that $\mathcal{D}({\rho'}({t_0}),{\texttt{R}_1}) > \mathcal{D}({\rho}({t_0}),{\texttt{R}_1})$. The QME occurs if there exist a time ${t_{\text{QME}}}$ such that for all $t > {t_{\text{QME}}}$ one finds $\mathcal{D}({e^{t\mathcal{L}}}[{\rho'}({t_0})],{\texttt{R}_1}) < \mathcal{D}({e^{t\mathcal{L}}}[\rho({t_0})],{\texttt{R}_1})$. (b) The protocol to accelerate convergence to equilibrium involves the following steps: (1) The unitary matrix $\Lambda$ maps the probe state to its diagonal form, (2) the permutation matrix ${P_{\pi}}$ rearranges the diagonal entries of ${\Lambda^{\dagger}}\rho({t_0}){\Lambda}$ so that the transformed state is as far away from equilibrium as possible, and (3) the resulting state from the previous step is rotated by the unitary matrix $U_1$ to the eigenbasis of the Hamiltonian.}
\label{fig:FIG00001}
\end{center}
\end{figure}

The Mpemba effect has found an analog in the quantum regime~\cite{PhysRevB.100.125102}. The so-called quantum Mpemba effect (QME) is defined as a physical process where a system initially further from equilibrium relaxes faster than a system closer to equilibrium. In Markovian open quantum systems, by acting a suitable unitary operator on the pure initial state of the system, one verifies the suppression of the slowest decay mode of the nonunitary dynamics~\cite{PhysRevLett.127.060401,PhysRevA.106.012207,g94p-7421}. This mechanism ensures an exponentially faster convergence to the stationary state, which defines the QME. It is worth noting, however, that the slowest decaying mode does not always contribute to the relaxation mechanism. This was observed, for example, in the effective Markovian dynamics of a single-level quantum dot coupled to two reservoirs~\cite{PhysRevLett.131.080402}. We refer the reader to Refs.~\cite{NatRevPhys_7_451_2025,arXiv:2502.01758,AAPPS_Bull_35_17_2025} for recent reviews on the QME.

It turns out that eliminating the slowest decaying mode is not enough to ensure the Mpemba effect. This issue motivated the proposal of a genuine quantum Mpemba effect, which involves accelerated relaxation to equilibrium and crossover between relaxation curves related to useful figures of merit that capture the distance from the probe state to the steady state~\cite{PhysRevLett.133.140404}. This setting involves mixed initial states and Davies maps whose spectral gap is dictated by a complex pair of eigenvalues related to a non-Hermitian jump operator. We highlight a recent study addressing the QME for open nonequilibrium quantum systems coupled to two different reservoirs~\cite{PhysRevLett.133.136302,PhysRevResearch.6.033330}, which paved the way for investigations concerning the so-called Pontus-Mpemba effect~\cite{hhgj-89gj,arXiv:2509.09366}.

The quantum Mpemba effect also manifests in the restoration of broken symmetries in many-body quantum systems. In this setting, the QME occurs when the symmetry is locally restored faster for the initial state which breaks it more~\cite{PhysRevLett.133.010401,Ares2023,PhysRevLett.133.140405,PhysRevB.110.085126,PhysRevB.111.104312,PhysRevA.111.043304,kppn-3272,arXiv:2507.16976}. We also mention recent works addressing the QME in the study of quantum complexity~\cite{arXiv:2509.22176,arXiv:2511.14837}, local relaxation of closed quantum systems~\cite{JPhysAMathTheor_58_445302_2025}, and quantum speed limits~\cite{PhysRevE.111.014133}. Experimental evidence of the quantum Mpemba effect was reported in trapped-ion quantum simulator~\cite{PhysRevLett.133.010402,NatCommun_16_301_2025}, including its inverse version~\cite{PhysRevLett.133.010403}, and more recently the QME was addressed in a nuclear magnetic resonance platform~\cite{arXiv:2511.14552}.

In this work, we present a framework that allows for exponentially faster relaxation to equilibrium and the QME in open quantum systems described by Davies maps [see Fig.~\ref{fig:FIG00001}(a)]. Our approach takes into account a unitary operator based on permutation matrices which, once applied to the initial state of the system, rearranges its spectrum and induces the suppression of the slowest decaying mode of the dynamics [see Fig.~\ref{fig:FIG00001}(b)]. On the one hand, we realize that exponential speed up is achieved regardless of the permutation matrix. On the other hand, we prove that there always exists a permutation matrix that maximizes the distance from the initial state to equilibrium. This distance is characterized by paradigmatic distinguishability measures, namely, Hilbert-Schmidt distance, quantum relative entropy, and trace distance. Together, these features guarantee the occurrence of the genuine quantum Mpemba effect. We present numerical simulations to support our findings. We emphasize that our approach can also be suitably adapted to encompass other classes of nonunitary evolutions, in particular recovering the main result of Ref.~\cite{PhysRevLett.127.060401} [see Appendix~\ref{sec:0000000000A}]. Finally, we show that achieving exponentially fast evolution toward equilibrium does not necessarily require the initial state to be incoherent with respect to the eigenstates of the Hamiltonian.

The outline of the paper is as follows. In Sec.~\ref{sec:00000000002}, we revise the general theory of Davies maps, particularly addressing the spectral properties of the Liouvillian that governs the effective nounitary dynamics of the quantum system. In Sec.~\ref{sec:00000000003}, we present a framework that engineers an exponential speedup relaxation by eliminating the slowest decay modes of the Liouvillian. In Sec.~\ref{sec:00000000004}, we show that it is always possible to find a permutation matrix such that we have a dressed state far from the initial state. Hence, once our protocol induces quicker relaxation towards equilibrium, these two features trigger a genuine quantum Mpemba effect. In Sec.~\ref{sec:00000000005}, we illustrate our findings for a two-level system, and for the dynamics of the transverse-field Ising model and the XXZ model, each of these spin chains weakly coupled to a thermal bath. In Sec.~\ref{sec:00000000006}, we show that there are initial states for which an exponentially rapid relaxation to the equilibrium is already observed, but which are not incoherent with respect to the energy eigenbasis. In Sec.~\ref{sec:00000000007} we summarize our conclusions.


\section{Markovian quantum dynamics}
\label{sec:00000000002}

Here we consider the effective nonunitary dynamics described by Davies maps, that is, quantum dynamical semigroups that describe the relaxation to the equilibrium of a given $d$-dimensional quantum system weakly coupled to a thermal bath at temperature $T$~\cite{Davies_1979,Breuer_Petruccione}. The nonunitary dynamics is governed by the Gorini-Kossakowski-Sudarshan-Lindblad master equation $d\rho(t)/dt = \mathcal{L}[\rho(t)]$, with the Liouvillian operator given by
\begin{equation}
\label{eq:0000001}
\mathcal{L}[\bullet] = -i[H,\bullet] + {\sum_{s = 1}^2}\, {\sum_{\substack{n,m = 1 \\ m < n}}^d}\,{\texttt{D}_{nm}^{(s)}}[\bullet] ~, 
\end{equation}
where the $s$-th dissipator operator is written as
\begin{equation}
\label{eq:0000002}
{\texttt{D}_{nm}^{(s)}}[\bullet] = {L_{nm}^{(s)}}\bullet{L_{n m}^{(s)\dagger}} - \frac{1}{2}\{{L_{n m}^{(s)\dagger}}{L_{nm}^{(s)}},\bullet\} ~,
\end{equation}
where $m,n\in\{1,\ldots,d\}$, $[\bullet,\bullet]$ and $\{\bullet,\bullet\}$ define the commutator and anticommutator, respectively, and we set $\hbar = 1$ throughout the manuscript. Here $H = {\sum_{l = 1}^d}\,{\varepsilon_l}|{\psi_l}\rangle\langle{\psi_l}|$ is the Hamiltonian of the system, where ${\{{\varepsilon_l}\}_{l = 1,\ldots,d}}$ are the energies, and ${\{|{\psi_l}\rangle\}_{l = 1,\ldots,d}}$ is the complete set of eigenvectors. In turn, the jump operators are given by~\cite{PhysRevLett.133.140404}
\begin{equation}
\label{eq:0000003}
{L_{nm}^{(1)}} = \sqrt{\xi_{nm}} \, |{\psi_n}\rangle\langle{\psi_m}| ~,\quad {L_{nm}^{(2)}} = \sqrt{\chi_{nm}} \, |{\psi_m}\rangle\langle{\psi_n}| ~,
\end{equation}
for $m,n = \{1,\ldots,d\}$, with $m < n$, and
\begin{equation}
\label{eq:0000004}
{\chi_{nm}}  = \frac{\gamma}{{e^{({\varepsilon_m} - {\varepsilon_n})/{k_B}T}} \pm 1} ~,\quad {\xi_{nm}} = \gamma \mp {\chi_{nm}} ~,
\end{equation}
where $\gamma$ is the strength of the coupling between system and environment, ${k_B}$ is the Boltzmann constant, and the signs $\pm$ refer to the Fermi-Dirac ($+$) and Bose-Einstein ($-$) distribution functions. We note that Eq.~\eqref{eq:0000001} excludes contributions from jump operators with $m = n$, which correspond to pure dephasing processes that leave populations unchanged. Such terms are associated with the zero-frequency component of the reservoir spectral density~\cite{10.1063/1.5115323,npjQuantumInf_6_74_2020}. In this work, however, we restrict our analysis to dissipative processes involving finite-frequency transitions between distinct energy eigenstates ($m \neq n$), which are sufficient to capture the key features of the relaxation dynamics of the open quantum system.

The jump operators in Eq.~\eqref{eq:0000003} define one possible realization of a Davies generator, for which the coherent and dissipative parts commute~\cite{JMathPhys_19_1227_1978,PhysRevA.68.032105}. It can be proved that the thermal Gibbs state is a fixed point for the Liouvillian characterized by ${L_{nm}^{(1)}}$ and ${L_{nm}^{(2)}}$, provided the detailed balance condition is satisfied~\cite{PhysRevLett.133.140404}. This condition constrains the transition rates via the Kubo-Martin-Schwinger relation, but does not require all transitions to be present~\cite{RevModPhys.89.015001}. Transitions that are absent correspond to vanishing rates and therefore trivially satisfy detailed balance. It is worth noting that some physical systems can violate the detailed balance condition while still approaching equilibrium~\cite{PhysRevB.90.220301,PhysRevB.106.115106,PhysRevE.88.020101}. In particular, Ref.~\cite{PhysRevLett.131.040401} demonstrated that for a quantum system interacting with a thermal reservoir of ideal particles, the steady state becomes a thermal Gibbs state at long times of the nonunitary dynamics, even though the detailed balance condition is violated beyond the Born approximation. Furthermore, Ref.~\cite{JPhysChemLett_16_4066_2025} showed that violation of the detailed balance condition can lead to oscillatory and polynomial decay in the thermalization of a nondegenerate, multilevel open quantum system. In this work, we focus on Davies maps that satisfy the detailed balance condition, thereby restricting our results to this physical scenario.

The formal solution of the Markovian master equation is given by $\rho(t) = {e^{t\mathcal{L}}}[\rho({t_0})]$, where $\rho({t_0})$ denotes the initial state of the system at time $t_0$. This state can be expressed in terms of the spectral decomposition of the Liouvillian superoperator as
\begin{equation}
\label{eq:0000005}
\rho(t) = {\texttt{R}_1} + {\sum_{s = 2}^{d^2}} \, {e^{-t|\text{Re}({\lambda_s})|}}{e^{it\text{Im}({\lambda_s})}}\text{Tr}({\texttt{L}_s}\rho({t_0})){\texttt{R}_s} ~,
\end{equation}
where ${\lambda_s} = \text{Re}({\lambda_s}) + i\text{Im}({\lambda_s})$ denotes the $s$-th eigenvalue of $\mathcal{L}$. The corresponding left and right eigenmatrices ${\texttt{L}_s}$ and ${\texttt{R}_s}$, satisfy ${\mathcal{L}^{\dagger}}[{\texttt{L}_s}] = {\lambda_s}{\texttt{L}_s}$ and ${\mathcal{L}}[{\texttt{R}_s}] = {\lambda_s}{\texttt{R}_s}$, respectively, where ${\mathcal{L}^{\dagger}}[\bullet]$ denotes the adjoint map. The sets of left and right eigenmatrices form a biorthogonal basis for the space of operators, obeying $\text{Tr}({\texttt{L}_s^{\dagger}}{\texttt{R}_l}) = {\delta_{sl}}$ for all $s,l \in \{1,\ldots,{d^2}\}$. The Liouvillian spectrum can be partitioned into two subsets: one consisting of $d(d - 1)$ complex eigenvalues appearing in conjugate pairs $\lambda_s$ and ${\lambda^*_s}$, and another consisting of $d$ real eigenvalues satisfying ${\lambda^*_s} = {\lambda_s}$. In the latter case, the corresponding eigenmatrices are Hermitian, i.e., ${\texttt{L}_s^{\dagger}} = {\texttt{L}_s}$ and ${\texttt{R}_s^{\dagger}} = {\texttt{R}_s}$.

To ensure completely positive dynamics, the nonunitary map in Eq.~\eqref{eq:0000005} must satisfy two conditions. First, the Liouvillian superoperator must have complex eigenvalues with nonpositive real parts, i.e., $\text{Re}({\lambda_s}) \leq 0$ for all $s = \{1,\ldots,{d^2}\}$, which implies $\text{Re}({\lambda_s}) = - |\text{Re}({\lambda_s})|$~\cite{Prosen}. Second, to guarantee that the nonunitary map is trace preserving, i.e., $\text{Tr}[\rho(t)] = 1$ for all $t \geq 0$, one of the eigenvalues of $\mathcal{L}$ must be zero, which we denote by ${\lambda_1} = 0$~\cite{RevModPhys.89.015001}. Under these conditions, the system asymptotically relaxes to a unique steady state ${\rho_{\infty}} := {\lim_{t \rightarrow \infty}}\rho(t)$, given by the eigenmatrix $\texttt{R}_1$~\cite{JStatMech_2019_043202}. Consistency further requires that the corresponding left eigenmatrix satisfies ${\texttt{L}_1} = \mathds{1}$, i.e., it is the identity operator. These properties are already incorporated in Eq.~\eqref{eq:0000005}.

For our purposes, the eigenvalues of the Liouville superoperator are arranged in ascending order according to the absolute value of their real parts, such that $0 = {\lambda_1} < |\text{Re}({\lambda_2})| \leq |\text{Re}({\lambda_3})| \leq \ldots  |\text{Re}({\lambda_{d^2}})|$. Hence, the mode with eigenvalue ${\lambda_2}$ [see Eq.~\eqref{eq:0000005}], having the smallest nonzero real part, sets the longest relaxation time, given by ${t_2} = 1/{|\text{Re}({\lambda_2})|}$. The slowest decay mode controls the relaxation timescale towards the steady state. For long times, the dynamics is expected to be dominated by such a mode, unless its overlap with the initial state becomes negligible. In fact, an exponential speed up in relaxation towards the equilibrium state is expected whenever the constraints $\text{Tr}({\texttt{L}_2^{\dagger}}\rho({t_0})) = 0$ and $\text{Tr}({\texttt{L}_2}\rho({t_0})) = 0$ are satisfied~\cite{PhysRevLett.127.060401}. More generally, this speedup can be engineered by a unitary matrix $U$ such that the transformed state ${\rho'}({t_0}) = U\rho({t_0}){U^{\dagger}}$ has a vanishing overlap with the lowest decaying eigenmatrix ${\texttt{L}_2}$, namely,
\begin{equation}
\label{eq:0000006}
\text{Tr}({\texttt{L}_2^{\dagger}}U\rho({t_0}){U^{\dagger}}) = 0 ~,
\end{equation}
and
\begin{equation}
\label{eq:0000007}
\text{Tr}({\texttt{L}_2}U\rho({t_0}){U^{\dagger}}) = 0 ~.
\end{equation}
This guarantees that the slowest mode is effectively suppressed, thereby accelerating relaxation to equilibrium~\cite{PhysRevA.106.012207}. As a consequence, the relaxation is governed by the next eigenvalue, denoted by $\widetilde{\lambda}$, which is real, with an associated timescale $1/|\widetilde{\lambda}|$, with $|\text{Re}({\lambda_2})| \leq |\widetilde{\lambda}|$. This construction leads to an exponential enhancement in the convergence toward the steady state, $\mathcal{D}({\rho'}(t),{\texttt{R}_1}) \sim \exp(-t|\widetilde{\lambda}|)$ for a given distinguishability measure. Accordingly, the relaxation dynamics of ${\rho'}(t) = {e^{t\mathcal{L}}}[{\rho'}({t_0})]$ becomes exponentially faster than that of ${\rho}(t) = {e^{t\mathcal{L}}}[{\rho}({t_0})]$. For clarity, the terms {\it exponential speedup} and {\it exponentially accelerated} follow the standard terminology in the quantum Mpemba effect literature~\cite{PhysRevLett.127.060401,PhysRevLett.133.140404}, which we adopt here as appropriate for the present context. This terminology should not be interpreted in the complexity-theoretic sense employed in quantum computing. Specifically, within the theory of algorithms and computational complexity, an exponential speedup refers to an asymptotic improvement in the scaling of computational resources with problem size, such as a transition between exponential and polynomial complexity~\cite{doi:10.1137/S0097539796300921,PhysRevLett.91.257902,PhysRevLett.127.060503,z2jq-1rxp}. In turn, the exponential speedup discussed in the context of the quantum Mpemba effect concerns the physical relaxation dynamics of a system and therefore represents a dynamical enhancement of thermalization or equilibration rather than an improvement in algorithmic efficiency. More precisely, the phenomenon describes situations in which quantum effects give rise to a relaxation rate that is exponentially larger than a reference rate, or equivalently, to a relaxation time that is exponentially shorter than that associated with an alternative initial preparation or dynamical pathway~\cite{PhysRevLett.133.136302,NatRevPhys_7_451_2025}. In this context, the exponential enhancement arises from the dependence of characteristic relaxation times on the physical parameters governing the dynamics. Next, we will address the issue of accelerating the relaxation process. Here, we focus on open quantum systems described by Davies maps, but we emphasize that our approach can encompass other types of dynamical maps [see Appendix~\ref{sec:0000000000A}].


\section{Speeding up the relaxation process}
\label{sec:00000000003}

We now focus on investigating a protocol to accelerate the relaxation of the open quantum system towards equilibrium. For our purposes, we recast the Hamiltonian as $H = {U_1}\varepsilon{U_1^{\dagger}}$, where $\varepsilon = \text{diag}({\varepsilon_1},\ldots,{\varepsilon_d})$ is the diagonal matrix that contains its energies, with ${\varepsilon_l} > {\varepsilon_{l + 1}}$ for all $l = \{1,\ldots,d - 1\}$, while ${U_1}$ is the unitary matrix formed by the respective eigenvectors. Overall, for a given initial state, eliminating the slowest decaying mode requires preparing a state ${\rho'}({t_0}) = {U}\rho({t_0}){U^{\dagger}}$ that is orthogonal to the eigenmatrix ${\texttt{L}_2}$ of the Liouvillian, for a given unitary matrix $U$. To achieve the results in Eqs.~\eqref{eq:0000006} and~\eqref{eq:0000007}, we consider the unitary operator 
\begin{equation}
\label{eq:0000008}
U = {U_1}{P_{\pi}}{\Lambda^{\dagger}} ~.
\end{equation}
Here, $\Lambda$ is the unitary matrix formed by the eigenstates of the initial state $\rho({t_0})$, such that $\rho({t_0}) = {\Lambda}{D}{\Lambda^{\dagger}}$. The matrix $D = \text{diag}({q_1},\ldots,{q_d})$ is diagonal, with entries given by the eigenvalues of the probe state. In turn, ${P_{\pi}}$ is a permutation matrix that reorders the spectrum of $\rho({t_0})$ in descending order such that ${P_{\pi}}{D}{P_{\pi}^{\dagger}} = \text{diag}({q_{\pi(1)}},\ldots,{q_{\pi(d)}})$, with ${q_{\pi(1)}} > \ldots > {q_{\pi(d)}}$, where $\pi(\bullet)$ denotes a permutation mapping each index $l \in \{1,\ldots,d\}$ to $\pi(l) \in \{1,\ldots,d\}$. By applying this unitary to the probe state, we obtain the density matrix
\begin{align}
\label{eq:0000009}
{\rho'}({t_0}) &= {U}\rho({t_0}){U^{\dagger}} \nonumber\\
&= {U_1}{P_{\pi}}{\Lambda^{\dagger}}\rho({t_0})\Lambda{P_{\pi}^{\dagger}}{U_1^{\dagger}} \nonumber\\
&= {U_1}{P_{\pi}}{D}{P_{\pi}^{\dagger}}{U_1^{\dagger}} ~,
\end{align}
which from now on we refer to a dressed initial state. In Fig.~\ref{fig:FIG00001}(b), we illustrate the role of the unitary matrix $U$. In Appendix~\ref{sec:0000000000A}, we show that Eq.~\eqref{eq:0000008} can be reformulated to recover the results discussed in Ref.~\cite{PhysRevLett.127.060401}.

The unitary transformation in Eq.~\eqref{eq:0000009} first maps $\rho({t_0})$ to its diagonal form. The matrix ${P_{\pi}}$ then reorders the diagonal entries of $D$, yielding ${P_{\pi}}{D}{P_{\pi}^{\dagger}} = \text{diag}({q_{\pi(1)}},\ldots,{q_{\pi(d)}})$, where the eigenvalues satisfy ${q_{\pi(1)}} > \ldots > {q_{\pi(d)}}$, i.e., they are sorted in descending order. Notably, ${P_{\pi}}$ also ensures that the transformed state ${\rho'}({t_0})$ lies further from the steady state than the original state ${\rho}({t_0})$, with respect to a given distinguishability measure on the space of quantum states. As will become clear later, this feature is essential for the emergence of a genuine Mpemba effect under a given nonunitary dynamics [see Sec.~\ref{sec:00000000004}]. Finally, the resulting matrix is rotated by $U_1$ to produce the quantum state in Eq.~\eqref{eq:0000009}, which is incoherent with respect to the eigenbasis of the Hamiltonian $H$. Compared with the results of Ref.~\cite{PhysRevLett.133.140404}, our protocol requires relatively low computational effort. In particular, the calculations rely on prior knowledge of the spectra of $H$ and $\rho({t_0})$. We emphasize that $P_{\pi}$ reorders the spectrum of $\rho({t_0})$ in descending order, provided the eigenvalues of $H$ are likewise sorted. For low-dimensional systems, $P_{\pi}$ can be constructed straightforwardly by permuting the rows and columns of $D$. Although larger systems, such as many-body quantum systems, may entail some computational cost overhead, no numerical optimization will be required to construct $P_{\pi}$.

The unitary in Eq.~\eqref{eq:0000008} causes the slowest decaying mode of the Liouvillian to be mapped onto the matrix ${\texttt{L}'_2} = {U_1^{\dagger}}{\texttt{L}_2}{U_1}$, where ${\texttt{L}'_2}$ is an upper or lower triangular matrix in the computational basis that has all elements zero except for one of its off-diagonal entries. In detail, we have that ${\texttt{L}'_2} = |{j_0}\rangle\langle{l_0}|$ for a given pair of states $|{j_0}\rangle$ and $|{l_0}\rangle$, with ${j_0} \neq {l_0}$, where ${\{|{j}\rangle\}_{j = 1,\ldots,d}}$ defines the computational basis related to the $d$-dimensional Hilbert space of the system [e.g., the set of eigenstates of the observable ${S_z} = (1/2){\sum_{j = 1}^d}{\sigma_j^z}$]. In this setting, given that ${U}\rho({t_0}){U^{\dagger}} = {U_1}{P_{\pi}}{D}{P^{\dagger}_{\pi}}{U_1^{\dagger}}$, we conclude that
\begin{align}
\label{eq:0000010}
\text{Tr}({\texttt{L}_2}{U}\rho({t_0}){U^{\dagger}}) &= \text{Tr}({\texttt{L}'_2}{P_{\pi}}{D}{P_{\pi}^{\dagger}}) \nonumber\\
&= \langle{l_0}|{P_{\pi}}{D}{P_{\pi}^{\dagger}}|{j_0}\rangle \nonumber\\
&= 0 ~,
\end{align}
where we used the fact that the permutation matrix simply rearranges the diagonal entries of the matrix $D$. The result in Eq.~\eqref{eq:0000010} holds for any permutation matrix ${P_{\pi}}$ (or even for a composition of permutation matrices). This approach allows the elimination of all $d(d - 1)$ eigenmatrices associated with complex eigenvalues of the Liouvillian that can be triangularized under the unitary transformation ${\texttt{L}_s} = {U_1}{\texttt{L}'_s}{U_1^{\dagger}}$. In other words, the framework is not limited to the eigenmatrix ${\texttt{L}_2}$. Instead, it is sufficiently robust to suppress additional excited modes that could otherwise hinder the exponential acceleration of the relaxation dynamics.

Below we present the proof that supports our claims. For our purposes, from now on we recast the eigenstates of the Hamiltonian to be written as $|{\psi_j}\rangle = {U_1}|{j}\rangle$ for all $j = \{1,\ldots,d\}$, where ${\{|{j}\rangle\}_{j = 1,\ldots,d}}$ defines the computational basis. In this case, the jump operators in Eq.~\eqref{eq:0000003} can be conveniently written as ${L_{n m}^{(1)}} = {\sqrt{\xi_{n m}}} \, {U_1}|{n}\rangle\langle{m}|{U_1^{\dagger}}$ and ${L_{n m}^{(2)}} = \sqrt{\chi_{n m}} \, {U_1}|{m}\rangle\langle{n}|{U_1^{\dagger}}$. The Liouvillian given in Eq.~\eqref{eq:0000001} is then vectorized by the Choi-Jamio{\l}kowski isomorphism~\cite{RepMathPhys_3_275_1972,LinAlgAppl_10_285_1975}, which yields
\begin{equation}
\label{eq:0000011}
{\mathcal{L}_{\text{vec}}} = ({U^*_1}\otimes{U_1})\, \Xi \,({U^{\top}_1}\otimes{U^{\dagger}_1}) ~,
\end{equation}
with
\begin{equation}
\label{eq:0000012}
\Xi := {A^{\dagger}}\otimes{I} + {I}\otimes{A} + {B} ~,
\end{equation}
where $A$ and $B$ are the auxiliary matrices given by
\begin{equation}
\label{eq:0000013}
A = - \frac{1}{2}\,{\sum_{\substack{n,m = 1 \\ m < n}}^d}\left({\xi_{nm}}|{m}\rangle\langle{m}| + {\chi_{nm}} |{n}\rangle\langle{n}|\right) - i \varepsilon
\end{equation}
and
\begin{equation}
\label{eq:0000014}
 B = {\sum_{\substack{n,m = 1 \\ m < n}}^d}\left({\chi_{nm}}|{m,m}\rangle\langle{n,n}| + {\xi_{nm}}|{n,n}\rangle\langle{m,m}|\right) ~,
\end{equation}
with $|m,n\rangle = |{m}\rangle\otimes|{n}\rangle$. Here ${A^{\dagger}}\otimes{I} + {I}\otimes{A}$ is a diagonal matrix, while $B$ is a non-diagonal matrix that has $d(d - 1)$ rows and columns with zero elements, since $m < n$. 

The operator $\Xi$ in Eq.~\eqref{eq:0000012} can be recast into a block-diagonal form $\Xi = {\Xi_{\text{diag}}}\oplus\,{\Xi_{\text{off}}}$ with respect  to the basis $\{|m,n\rangle\}_{m,n = 1,\ldots,d}$, where ${\Xi_{\text{diag}}}$ is a diagonal matrix with $d(d - 1)$ rows and columns, while ${\Xi_{\text{off}}}$ is an off-diagonal matrix with $d$ rows and columns. In addition, it has a spectral decomposition given by $\Xi = {\sum_k}\,{\lambda_k}|{\texttt{R}'_k}\rangle\rangle\langle\langle{\texttt{L}'_k}|$, where left $|{\texttt{L}'_k}\rangle\rangle$ and right $|{\texttt{R}'_k}\rangle\rangle$ eigenvectors are biorthogonal to each other as $\langle\langle{\texttt{L}'_j}|{\texttt{R}'_k}\rangle\rangle = {\delta_{jk}}$, for all $j,k = \{1,\ldots,{d^2}\}$. In this case, it follows that the spectral decomposition of the Liouvillian in Eq.~\eqref{eq:0000011} is written as ${\mathcal{L}_{\text{vec}}} = {\sum_k}\,{\lambda_k}|{\texttt{R}_k}\rangle\rangle\langle\langle{\texttt{L}_k}|$, where we define $|{X_k}\rangle\rangle := ({U^*_1}\otimes{U_1})|{X'_k}\rangle\rangle$, for all $X \in \{\texttt{L},\texttt{R}\}$. We emphasize that $|{X'_k}\rangle\rangle$ and $|{X_k}\rangle\rangle$ stand for the vector forms of eigenmatrices $X'_k$ and ${X_k} = {U_1}{X'_k}{U_1^{\dagger}}$, respectively. In addition, we have that vectors $|{\texttt{L}_j}\rangle\rangle$ and $|{\texttt{R}_j}\rangle\rangle$ satisfy the or\-tho\-go\-na\-li\-ty constraint $\langle\langle{\texttt{L}_j}|{\texttt{R}_l}\rangle\rangle = \langle\langle{\texttt{L}'_j}|({U_1^{\top}}\otimes{U^{\dagger}_1})({U^*_1}\otimes{U_1})|{\texttt{R}'_l}\rangle\rangle = \langle\langle{\texttt{L}'_j}|{\texttt{R}'_l}\rangle\rangle = {\delta_{jl}}$.

We observe that the operators ${\mathcal{L}_{\text{vec}}}$ and $\Xi$ share the same spectrum of ${d^2}$ eigenvalues, of which $d$ are real and come from the subblock ${\Xi_{\text{off}}}$, and the remaining $d(d - 1)$ eigenvalues are complex and related to ${\Xi_{\text{diag}}}$. On the one hand, we find that the subblock ${\Xi_{\text{off}}}$ contributes with $d$ eigenvectors $|{\texttt{L}'_s}\rangle\rangle$ and $|{\texttt{R}'_s}\rangle\rangle$, whose matrix forms are Hermitian, fully diagonal eigenmatrices ${\texttt{L}'_s}$ and ${\texttt{R}'_s}$ with respect to the computational basis $\{|{k}\rangle\}_{k = 1,\ldots,d}$. This naturally includes the eigenoperator ${\texttt{R}'_1}$ related to the steady state ${\texttt{R}_1} = {U_1}{\texttt{R}'_1}{U_1^{\dagger}}$ of the dynamics and also the vector $|{\texttt{L}'_1}\rangle\rangle$ constrained to the eigenmatrix ${\texttt{L}_1}$ that plays the role of the identity. On the other hand, the slowest decaying modes of the nonunitary dynamics are related to the subspace spanned by $d(d - 1)$ eigenvectors of ${\Xi_{\text{diag}}}$. Note that the respective set of left $|{\texttt{L}'_s}\rangle\rangle$ and right $|{\texttt{R}'_s}\rangle\rangle$ eigenvectors gives rise to upper and lower triangular matrices $\texttt{L}'_k$ and ${\texttt{R}'_s}$ in the computational basis, respectively, which have all entries equal to zero except for one off-diagonal element. In detail, one finds that $\langle{j}|{\texttt{L}'_s}|{l}\rangle = {\delta_{j,{j_0}}}{\delta_{l,{l_0}}}$ and $\langle{j}|{\texttt{R}'_s}|{l}\rangle = {\delta_{j,{j_0}}}{\delta_{l,{l_0}}}$, for a given pair ${j_0},{l_0} \in \{1,\ldots,d\}$, with ${j_0} \neq {l_0}$. In par\-ti\-cular, we note that this set comprises the eigenoperator ${\texttt{L}'_2} = {U_1^{\dagger}}{\texttt{L}_2}{U_1}$ connected to the lowest decaying mode of the Davies map. It turns out that ${\texttt{L}'_2} = |{j_0}\rangle\langle{l_0}|$ is a non-Hermitian upper or lower triangular matrix. Finally, we note that, for higher-dimensional systems, the spectral analysis of the Liouvillian can be implemented using the results presented in Ref.~\cite{arXiv:2509.07709}.


\section{Quantum Mpemba effect}
\label{sec:00000000004}

In general, suppressing the slowest decaying modes of the dynamics ensures an exponential speedup in the relaxation of the open quantum system. However, this is not a sufficient criterion for the occurrence of a genuine quantum Mpemba effect. The general idea is to identify situations in which a quantum system initially further from equilibrium approaches it more rapidly than an initial state closer to it. To this end, an information-theoretic measure $\mathcal{D}(x,y)$ is introduced to monitor the nonunitary dynamics of the quantum system, thus investigating the distinguishability of the instantaneous state $\rho(t)$ and the steady state $\texttt{R}_1$. The genuine QME is evidenced by the crossover between relaxation curves related to initial and transformed states. Useful figures of merit include the Hilbert-Schmidt distance~\cite{PhysRevLett.127.060401}, the quantum relative entropy~\cite{PhysRevLett.133.140404}, and the trace distance~\cite{arXiv:2511.04353}, to name a few. 

Here we address the quantum Mpemba effect for open quantum systems whose effective dynamics is governed by Davies maps. Our main result is that the framework described in Sec.~\ref{sec:00000000003} naturally contributes to the occurrence of the QME, since (i) it eliminates slower decay modes to accelerate relaxation towards equilibrium, and (ii) it ensures that the dressed state ${\rho'}({t_0}) = {U}\rho({t_0}){U^{\dagger}}$ is as far away from equilibrium as possible compared to the initial state ${\rho}({t_0})$, where $U = {U_1}{P_{\pi}}{\Lambda^{\dagger}}$. This last requirement means that, for a given distingui\-sha\-bi\-li\-ty measure $\mathcal{D}(x,y)$, the distance from ${\rho'}({t_0})$ to the equilibrium is greater than the distance from $\rho({t_0})$ to the steady state ${\texttt{R}_1}$, that is,
\begin{equation}
\label{eq:0000015}
\mathcal{D}({\rho'}({t_0}),{\texttt{R}_1}) > \mathcal{D}(\rho({t_0}),{\texttt{R}_1}) ~.
\end{equation}
We note that ${\texttt{R}_1} = {U_1}\Sigma{U_1^{\dagger}}$ defines the steady state of the system, with $\Sigma = \exp(-\varepsilon/{k_B}T)/Z$, and $Z = \text{Tr}(\exp(-\varepsilon/{k_B}T))$ is the partition function. Once conditions (i) and (ii) are fully satisfied, a genuine QME occurs if, at some later time ${t_{\text{QME}}} > {t_0}$, the reverse bound
\begin{equation}
\label{eq:0000016}
\mathcal{D}({\rho'}(t),{\texttt{R}_1}) < \mathcal{D}(\rho(t),{\texttt{R}_1})
\end{equation}
is found for all $t > {t_{\text{QME}}}$, where ${\rho'}(t) = {e^{t\mathcal{L}}}[{\rho'}({t_0})]$ and $\rho(t) = {e^{t\mathcal{L}}}[{\rho}({t_0})]$ define the ins\-tan\-taneous states obtained from ${\rho'}({t_0})$ and $\rho({t_0})$, respectively, both evol\-ving under the generator in Eq.~\eqref{eq:0000001}. 

To be valid, the bound in Eq.~\eqref{eq:0000016} requires that the two curves of $\mathcal{D}({\rho'}(t),{\texttt{R}_1})$ and $\mathcal{D}(\rho(t),{\texttt{R}_1})$ intersect at time ${t_{\text{QME}}}$. This crossover remains as a consequence of points (i) and (ii) to be satisfied for the proposed unitary transformation. We have already established point (i), namely that the unitary $U$ contributes to accelerating relaxation. More precisely, the speedup arises because applying $U$ to the state $\rho({t_0})$ suppresses contributions from Liouvillian eigenmodes with complex eigenvalues, by removing their overlap with the dressed state. This mechanism is independent of the time required to prepare the transformed state ${\rho'}({t_0})$. In the following, we verify point (ii) by showing that the bound in Eq.~\eqref{eq:0000015} is valid for three paradigmatic distinguishability measures, namely, Hilbert-Schmidt distance, quantum relative entropy, and trace distance. In detail, we prove that there always exists a permutation matrix ${P_{\pi}}$ that maximizes the distance $\mathcal{D}({\rho'}({t_0}),{\texttt{R}_1})$ from the dressed state to the equilibrium, compared to $\mathcal{D}({\rho}({t_0}),{\texttt{R}_1})$. Some technical details of the proofs can be found in Appendix~\ref{sec:0000000000B}.


\subsection{Hilbert-Schmidt distance}
\label{sec:00000000004A}

Let $\rho \in \mathcal{S}$ be two density matrices defined on the convex space of quantum states $\mathcal{S} = \{\rho \in \mathcal{H}~|~\rho^{\dagger} = \rho,~\rho \geq 0,~\text{Tr}(\rho) = 1\}$, where $\mathcal{H}$ is a $d$-dimensional Hilbert space. The Hilbert-Schmidt distance (HSD) between these two states is defined as
\begin{equation}
\label{eq:0000017}
{\mathcal{D}_{\text{HSD}}}(\rho,\varrho) = \sqrt{\text{Tr}[{(\rho - \varrho)^2}]} ~.
\end{equation}
We note that the HSD is (i) non-negative, i.e., ${\mathcal{D}_{\text{HSD}}}(\rho,\varrho) \geq 0$, with ${\mathcal{D}_{\text{HSD}}}(\rho,\varrho) = 0$ if and only if $\rho = \varrho$; (ii) symmetric, ${\mathcal{D}_{\text{HSD}}}(\rho,\varrho) = {\mathcal{D}_{\text{HSD}}}(\varrho,\rho)$; and (iii) isometric invariant, ${\mathcal{D}_{\text{HSD}}}(V\rho{V^{\dagger}}, V\varrho{V^{\dagger}}) = {\mathcal{D}_{\text{HSD}}}(\rho,\varrho)$, where ${V^{\dagger}} = {V^{-1}}$ is a unitary operator. It is worth noting that the HSD sa\-tis\-fies the triangle inequality ${\mathcal{D}_{\text{HSD}}}(\rho,\varrho) \leq {\mathcal{D}_{\text{HSD}}}(\rho,\varpi) + {\mathcal{D}_{\text{HSD}}}(\varpi,\varrho)$, for all $\rho, \varpi, \varrho \in \mathcal{S}$. However, the HSD it is not always contractive under completely positive and trace-preserving maps~\cite{PhysLettA_268_158_2000}.

The bound in Eq.~\eqref{eq:0000015}, applied to the Hilbert-Schmidt distance, becomes ${\mathcal{D}_{\text{HSD}}}({\rho'}({t_0}),{\texttt{R}_1}) > {\mathcal{D}_{\text{HSD}}}({\rho}({t_0}),{\texttt{R}_1})$. This inequality, when satisfied, must imply the upper bound
\begin{equation}
\label{eq:0000018}
\text{Tr}({\rho'}({t_0}){\texttt{R}_1}) < \text{Tr}({\rho}({t_0}){\texttt{R}_1}) ~.
\end{equation}
In Appendix~\ref{sec:0000000000B}, we show that Eq.~\eqref{eq:0000018} can be written as
\begin{equation}
\label{eq:0000019}
\text{Tr}({P_{\pi}}{D}{P^{\dagger}_{\pi}}{\Sigma}) < {\sum_l}\,{\eta_l}\text{Tr}({A_l}{D}{A_l^{\dagger}}\Sigma) ~,
\end{equation}
where $A_l$ is a permutation matrix, $0 \leq {\eta_l} \leq 1$, with ${\sum_l}\,{\eta_l} = 1$, and $D = \text{diag}({q_1},\ldots,{q_d})$ and $\Sigma = \text{diag}({\alpha_1},\ldots,{\alpha_d})$ are diagonal matrices containing the spectrum of $\rho({t_0})$ and ${\texttt{R}_1}$, respectively. The right-hand side of Eq.~\eqref{eq:0000019} defines a convex sum ${\sum_l}\,{\eta_l}{c_l}$ of the nonnegative elements ${c_l} := \text{Tr}({A_l}{D}{A_l^{\dagger}}\Sigma)$. We note that ${c_{\text{min}}} \leq {\sum_l}\,{\eta_l}{c_l} \leq {c_{\text{max}}}$, where ${c_{\text{min}}} = \text{min}\{{c_1},\ldots,{c_d}\}$ and ${c_{\text{max}}} = \text{max}\{{c_1},\ldots,{c_d}\}$. This means that there exists an optimal permutation matrix ${P_{\pi,\text{opt}}}$ such that $\text{Tr}({P_{\pi,\text{opt}}}{D}{P_{\pi,\text{opt}}^{\dagger}}\Sigma)$ is closest to the minimum value $c_{\text{min}}$, i.e., $\text{Tr}({P_{\pi,\text{opt}}}{D}{P_{\pi,\text{opt}}^{\dagger}}\Sigma) \leq {c_{\text{min}}}$. To see this point, we note that 
\begin{equation}
\label{eq:0000020}
\text{Tr}({P_{\pi,\text{opt}}}{D}{P_{\pi,\text{opt}}^{\dagger}}\Sigma) = {\sum_{l = 1}^d}\,{\alpha_l}{q_{\pi(l)}} ~, 
\end{equation}
where $\pi(\bullet)$ maps index $l$ to $\pi(l)$, with $l,\pi(l) \in \{1,\ldots,d\}$. Therefore, to ensure that ${\sum_{l = 1}^d}\,{\alpha_l}{q_{\pi(l)}} \leq {c_{\text{min}}}$, it suffices to choose a permutation matrix such that ${P_{\pi,\text{opt}}}{D}{P^{\dagger}_{\pi,\text{opt}}} = \text{diag}({q_{\pi(1)}},\ldots,{q_{\pi(d)}})$, with ${q_{\pi(1)}} \geq \ldots \geq {q_{\pi(d)}}$ being listed in descending order provided ${\alpha_1} \leq \ldots \leq {\alpha_n}$ are listed in ascending order, or vice versa~\cite{hardy1952inequalities}. This proves the validity of the inequality in Eq.~\eqref{eq:0000018}, which means that there exists a permutation $P_{\pi}$ that maximizes the HSD evaluated for states ${\rho'}({t_0})$ and ${\texttt{R}_1}$.


\subsection{Quantum relative entropy}
\label{sec:00000000004B}

The quantum relative entropy (QRE) with respect to states $\rho,\varrho\in\mathcal{S}$ is defined as
\begin{equation}
\label{eq:0000021}
{\mathcal{D}_{\text{QRE}}}(\rho,\varrho) := - S(\rho) - \text{Tr}(\rho\ln\varrho) ~,
\end{equation}
where $S(\rho) = -\text{Tr}(\rho\ln\rho)$ is the von Neumann entropy~\cite{RelEntr01}. To be clear, the QRE is (i) non-negative, i.e., ${\mathcal{D}_{\text{QRE}}}(\rho,\varrho) \geq 0$, with ${\mathcal{D}_{\text{QRE}}}(\rho,\varrho) = 0$ if and only if $\rho = \varrho$; (ii) isometric invariant, ${\mathcal{D}_{\text{QRE}}}(V\rho{V^{\dagger}}, V\varrho{V^{\dagger}}) = {\mathcal{D}_{\text{QRE}}}(\rho,\varrho)$, where ${V^{\dagger}} = {V^{-1}}$ is a unitary operator; (iii) monotonically decreasing under completely positive and trace preserving (CPTP) maps, i.e., ${\mathcal{D}_{\text{QRE}}}\left(\mathcal{E}(\rho), \mathcal{E}(\varrho)\right) \leq {\mathcal{D}_{\text{QRE}}}(\rho, \varrho)$, with $\mathcal{E}(\bullet)$ a given CPTP operation~\cite{RevModPhys.74.197,arXiv:quant-ph_0004045}. However, the QRE does not define a {\it bona fide} metric, as it is not symmetric under the exchange of its arguments, i.e., ${\mathcal{D}_{\text{QRE}}}(\rho,\varrho) \neq {\mathcal{D}_{\text{QRE}}}(\varrho,\rho)$.

The bound in Eq.~\eqref{eq:0000016}, applied to the QRE, becomes ${\mathcal{D}_{\text{QRE}}}({\rho'}({t_0}),{\texttt{R}_1}) > {\mathcal{D}_{\text{QRE}}}({\rho}({t_0}),{\texttt{R}_1})$. This inequality, when satisfied, readily implies the lower bound
\begin{equation}
\label{eq:0000022}
\text{Tr}({\rho'}({t_0})H) > \text{Tr}({\rho}({t_0})H) ~,
\end{equation}
where we used the fact that the von Neumann entropy is unitarily invariant, i.e., $S({\rho'}({t_0})) = S(U\rho({t_0}){U^{\dagger}}) = S({\rho}({t_0}))$. We also used that $\text{Tr}(X\ln{\texttt{R}_1}) = - \beta\text{Tr}(XH) - \ln{Z}$, which holds for $X = \rho({t_0})$, and $X = {\rho'}({t_0})$. Next we recall that ${\rho'}({t_0}) = {U_1}{P_{\pi}}{D}{P_{\pi}^{\dagger}}{U_1^{\dagger}}$ and $H = {U_1}\varepsilon{U_1^{\dagger}}$, so that 
\begin{equation}
\label{eq:0000023}
\text{Tr}({\rho'}({t_0})H) = \text{Tr}({P_{\pi}}D{P_{\pi}^{\dagger}}\varepsilon) ~.
\end{equation}
Based on the results from Appendix~\ref{sec:0000000000B}, it can be verify that 
\begin{equation}
\label{eq:0000024}
\text{Tr}({\rho}({t_0})H) = {\sum_l}\,{\eta_l}\text{Tr}({A_l}{D}{A_l^{\dagger}}\varepsilon) ~, 
\end{equation}
where $A_l$ is a permutation matrix, and $0 \leq {\eta_l} \leq 1$, with ${\sum_l}\,{\eta_l} = 1$. Hence, combining Eqs.~\eqref{eq:0000022},~\eqref{eq:0000023}, and~\eqref{eq:0000024} yields
\begin{equation}
\label{eq:0000025}
\text{Tr}({P_{\pi}}{D}{P_{\pi}^{\dagger}}\varepsilon) > {\sum_l}\,{\eta_l}\text{Tr}({A_l}{D}{A_l^{\dagger}}\varepsilon) ~.
\end{equation}
To validate the inequality in Eq.~\eqref{eq:0000025}, we apply the same reasoning as in Sec.~\ref{sec:00000000004B}. The right-hand side of Eq.~\eqref{eq:0000025} defines a convex sum ${\sum_l}\,{\eta_l}{g_l}$ of the nonnegative elements ${g_l} := \text{Tr}({A_l}{D}{A_l^{\dagger}}\varepsilon)$. The idea is that there exists an optimal permutation matrix ${P_{\pi,\text{opt}}}$ that sa\-tis\-fies the lower bound $\text{Tr}({P_{\pi,\text{opt}}}{D}{P_{\pi,\text{opt}}^{\dagger}}\varepsilon) \geq {g_{\text{max}}} \geq {\sum_l}\,{\eta_l}{g_l}$, where ${g_{\text{max}}} = \text{max}\{{g_1},\ldots,{g_d}\}$. We note that $\text{Tr}({P_{\pi,\text{opt}}}{D}{P_{\pi,\text{opt}}^{\dagger}}\varepsilon) = {\sum_l}\,{\varepsilon_l}{q_{\pi(l)}}$. Therefore, to ensure that ${\sum_l}\,{\varepsilon_l}{q_{\pi(l)}} \geq {g_{\text{max}}} \geq {\sum_l}\,{\eta_l}{g_l}$, it is sufficient to choose a permutation matrix such that ${P_{\pi,\text{opt}}}{D}{P^{\dagger}_{\pi,\text{opt}}} = \text{diag}({q_{\pi(1)}},\ldots,{q_{\pi(d)}})$, with ${q_{\pi(1)}} \geq \ldots \geq {q_{\pi(d)}}$ listed in descending order provided ${\varepsilon_1} \geq \ldots \geq {\varepsilon_n}$ are listed in descending order, or vice versa~\cite{hardy1952inequalities}. This proves the inequality in Eq.~\eqref{eq:0000022}, which means that there exists $P_{\pi}$ that maximizes the QRE evaluated for ${\rho'}({t_0})$ and ${\texttt{R}_1}$.


\subsection{Trace distance}
\label{sec:00000000004C}

The trace distance (TD) for two quantum states $\rho,\varrho\in\mathcal{S}$ is defined as
\begin{equation}
\label{eq:0000026}
{\mathcal{D}_{\text{TD}}}(\rho,\varrho) = \frac{1}{2}\text{Tr}(|\rho - \varrho|) ~,
\end{equation}
with $|X| := \sqrt{{X^{\dagger}}X}$. Overall, the TD is (i) non-negative, i.e., ${\mathcal{D}_{\text{TD}}}(\rho,\varrho) \geq 0$, with ${\mathcal{D}_{\text{TD}}}(\rho,\varrho) = 0$ if and only if $\rho = \varrho$; (ii) symmetric, ${\mathcal{D}_{\text{TD}}}(\rho,\varrho) = {\mathcal{D}_{\text{TD}}}(\varrho,\rho)$; (iii) unitarily invariant, ${\mathcal{D}_{\text{TD}}}(V\rho{V^{\dagger}}, V\varrho{V^{\dagger}}) = {\mathcal{D}_{\text{TD}}}(\rho,\varrho)$, where ${V^{\dagger}} = {V^{-1}}$ is a unitary operator; and (iv) monotonically decreasing under CPTP maps, i.e., ${\mathcal{D}_{\text{TD}}}\left(\mathcal{E}(\rho), \mathcal{E}(\varrho)\right) \leq  {\mathcal{D}_{\text{TD}}}(\rho, \varrho)$, with $\mathcal{E}(\bullet)$ a given CPTP operation~\cite{Nielsen_Chuang_infor_geom}.

To investigate the QME, we first note that the trace distance for the initial state $\rho({t_0})$ and the steady state $\texttt{R}_1$ satisfies the lower bound~\cite{PhysRevA.77.042111} 
\begin{equation}
\label{eq:0000027}
{\mathcal{D}_{\text{TD}}}({u^{\uparrow}},{v^{\downarrow}}) \geq {\mathcal{D}_{\text{TD}}}(\rho({t_0}),{\texttt{R}_1}) ~,
\end{equation}
where ${u^{\uparrow}} = \text{diag}({q_1},\ldots,{q_d})$ defines a matrix formed by the eigenvalues of $\rho({t_0})$ listed in ascending order, while ${v^{\downarrow}} = \text{diag}({\alpha_1},\ldots,{\alpha_d})$ is a matrix that contains the eigenvalues of $\texttt{R}_1$ listed in descending order, with ${\alpha_l} = e^{-{\varepsilon_l}/{k_B}T}/Z$ the $l$-th eigenvalue of the steady state. Next, note that the trace distance for the initial dressed state ${\rho'}({t_0})$ and the steady state $\texttt{R}_1$ is written as
\begin{align}
\label{eq:0000028}
{\mathcal{D}_{\text{TD}}}({\rho'}({t_0}),{\texttt{R}_1}) &= \frac{1}{2}\text{Tr}\left[\sqrt{({U_1}{P_{\pi}}{D}{P_{\pi}^{\dagger}}{U_1^{\dagger}} - {U_1}\Sigma {U_1^{\dagger}})^2} \, \right] \nonumber\\
& = \frac{1}{2}\text{Tr}\left[\sqrt{({P_{\pi}}{D}{P_{\pi}^{\dagger}} - \Sigma)^2} \, \right] \nonumber\\
& = {\mathcal{D}_{\text{TD}}}({P_{\pi}}{D}{P_{\pi}^{\dagger}},\Sigma) ~,
\end{align}
where we used the fact the TD is unitarily invariant. Here $\Sigma = \text{diag}({\alpha_1},\ldots,{\alpha_d})$, while ${P_{\pi}}{D}{P_{\pi}^{\dagger}} = \text{diag}({q_{\pi(1)}},\ldots,{q_{\pi(d)}})$ is the matrix obtained by rear\-ran\-ging the entries of the diagonal matrix $D$ that contains the eigenvalues of the initial state. It is worth noting that there always exists an appropriate permutation matrix ${P_{\pi}}$ that rearranges the eigenvalues of the initial state such that ${\mathcal{D}_{\text{TD}}}({P_{\pi}}{D}{P_{\pi}^{\dagger}},\Sigma) = {\mathcal{D}_{\text{TD}}}({u^{\uparrow}},{v^{\downarrow}})$. In this case, combining Eqs.~\eqref{eq:0000027} and~\eqref{eq:0000028}, yields
\begin{equation}
\label{eq:0000029}
{\mathcal{D}_{\text{TD}}}({\rho'}({t_0}),{\texttt{R}_1}) \geq {\mathcal{D}_{\text{Tr}}}(\rho({t_0}),{\texttt{R}_1}) ~.
\end{equation}
Hence, having satisfied the inequality in Eq.~\eqref{eq:0000029}, that is, when the trace distance between the dressed initial state and the steady state is maximized for a given permutation matrix, a genuine quantum Mpemba effect is expected to be observed.


\section{Examples}
\label{sec:00000000005}

In this section we illustrate our findings for the case of the dynamics of (i) two-level systems [see Sec.~\ref{sec:00000000005A}]; and (ii) many-body quantum systems [see Sec.~\ref{sec:00000000005B}], particularly addressing the transverse-field Ising model and also the XXZ model.


\subsection{Two-level system}
\label{sec:00000000005A}

We consider a two-level system ($d = 2$) weakly coupled to a bosonic reservoir at temperature $T$. The Hamiltonian of the system is given by $H = {U_1}{\varepsilon}{U_1^{\dagger}}$, where ${\varepsilon} = \text{diag}({\varepsilon_1},{\varepsilon_2})$, with ${\varepsilon_1} > {\varepsilon_2}$, while ${U_1} = i(|{1}\rangle\langle{0}| - |{0}\rangle\langle{1}|)$. The vectors $\{|{0}\rangle,|{1}\rangle\}$ are the eigenstates of Pauli matrix ${\sigma_z}$, i.e., ${\sigma_z}|{x}\rangle = {(-1)^x}|{x}\rangle$ for $x = \{0,1\}$, and define the computational basis. The system is initialized at ${t_0} = 0$ in the pure single-qubit state $\rho({t_0}) = {\Lambda}{D}{\Lambda^{\dagger}}$, where $D = \text{diag}(0,1) = |1\rangle\langle{1}|$ and $\Lambda = {-i}|{0}\rangle\langle{-}| + |{1}\rangle\langle{+}|$, with $|{\pm}\rangle = (1/\sqrt{2})(|{0}\rangle \pm |{1}\rangle)$. The effective dynamics of the two-level system is go\-ver\-ned by the Liouvillian $\mathcal{L}[\bullet]$ in Eq.~\eqref{eq:0000001} [see also Eq.~\eqref{eq:0000002}], which in turn is related to the Lindblad operators [see Eq.~\eqref{eq:0000003}]
\begin{align}
\label{eq:0000030}
{L^{(1)}_{21}} &= {e^{\frac{\delta}{2{k_B}T}}}{({L^{(2)}_{21}})^{\dagger}} \nonumber\\
&= {e^{\frac{\delta}{4{k_B}T}}}\sqrt{\frac{\gamma}{2}\text{csch}\left(\frac{\delta}{2{k_B}T}\right)} |{0}\rangle\langle{1}| ~,
\end{align}
where $\delta = {\varepsilon_1} - {\varepsilon_2}$ is the gap of the two-level system, and $\gamma$ is the strength of the coupling between system and environment. 

The spectral decomposition of the Liouvillian is given in Table~\ref{tab:TABLE01}, with left eigenmatrices ${\texttt{L}_s} = {U_1}{\texttt{L}'_s}{U_1^{\dagger}}$ and right eigenmatrices ${\texttt{R}_s} = {U_1}{\texttt{R}'_s}{U_1^{\dagger}}$, for $s \in \{1,\ldots,4\}$. The slowest decaying mode is related to the complex pair of eigenvalues ${\lambda_2}$ and ${\lambda_3} = {\lambda^*_2}$. In particular, we ve\-ri\-fy that ${\texttt{L}_1} = |0\rangle\langle{0}| + |1\rangle\langle{1}|$ is the identity matrix, while ${\texttt{R}_1} = {p_+}|0\rangle\langle{0}| + {p_-}|1\rangle\langle{1}|$ is the steady state of the dy\-na\-mics, with ${p_{\mp}} = (1/2){e^{\mp\delta/(2{k_B}T)}}\text{sech}(\delta/(2{k_B}T))$. It can be proved that ${p_-} = {e^{-{\varepsilon_1}/({k_B}T)}}/Z$ and ${p_+} = {e^{-{\varepsilon_2}/({k_B}T)}}/Z$ are Boltzmann weights related to a thermal state at finite temperature $T$, with $Z = \text{Tr}({e^{-H/{k_B}T}})$ the partition function.
\begin{table}[!t]
\caption{Spectral decomposition of the Liouvillian operator that governs the effective nonunitary dynamics of the two-level system described in Sec.~\ref{sec:00000000005A}. We recall that ${\texttt{L}_s} = {U_1}{\texttt{L}'_s}{U_1^{\dagger}}$ and ${\texttt{R}_s} = {U_1}{\texttt{R}'_s}{U_1^{\dagger}}$, for $s \in \{1,\ldots,4\}$, with ${U_1} = i(|{1}\rangle\langle{0}| - |{0}\rangle\langle{1}|)$. Here we introduce the Boltzmann weights ${p_{\mp}} = (1/2){e^{\mp\delta/(2{k_B}T)}}\text{sech}(\delta/(2{k_B}T))$, where $\delta = {\varepsilon_1} - {\varepsilon_2}$ is the gap of the two-level system, $\gamma$ is the strength of the coupling between system and environment, $T$ is the temperature of the bosonic reservoir, and $k_B$ is the Boltzmann constant.}
\begin{center}
\begin{tabular}{cc}
\hline\hline
Eigenvalue & Eigenmatrix \\
\hline
\multirow{2}{8em}{\[{\lambda_1} = 0\]} & ${\texttt{L}'_1} = |{0}\rangle\langle{0}| + |{1}\rangle\langle{1}|$ \\
& ${\texttt{R}'_1} = {p_-}|{0}\rangle\langle{0}| + {p_+}|{1}\rangle\langle{1}|$ \\
\hline
\multirow{2}{8em}{$\begin{aligned}&{\lambda_2} = \\ &- \frac{\gamma}{2}\coth\left(\frac{\delta}{2{k_B}T}\right) - i\delta\end{aligned}$} & ${\texttt{L}'_2} = |{0}\rangle\langle{1}|$  \\
& ${\texttt{R}'_2} = |{0}\rangle\langle{1}|$  \\
\hline
\multirow{2}{8em}{$\begin{aligned}&{\lambda_3} = {\lambda^*_2} = \\ &- \frac{\gamma}{2}\coth\left(\frac{\delta}{2{k_B}T}\right) + i\delta\end{aligned}$} & ${\texttt{L}'_3} = {\texttt{L}'^{\dagger}_2} = |{1}\rangle\langle{0}|$  \\
& ${\texttt{R}'_3} = {\texttt{R}'^{\dagger}_2} = |{1}\rangle\langle{0}|$  \\
\hline
\multirow{2}{8em}{${\lambda_4} = - \gamma\coth\left(\frac{\delta}{2{k_B}T}\right)$} & ${\texttt{L}'_4} = {p_-}|1\rangle\langle{1}| - {p_+}|0\rangle\langle{0}|$ \\
& ${\texttt{R}'_4} = |{1}\rangle\langle{1}| - |{0}\rangle\langle{0}|$ \\
\hline\hline
\end{tabular}
\label{tab:TABLE01}
\end{center}
\end{table}

In this setting, since $\text{Tr}({\texttt{L}_2}\rho({t_0})) \neq 0$ and $\text{Tr}({\texttt{L}_3}\rho({t_0})) \neq 0$, we apply our protocol to accelerate relaxation to equilibrium and also induce the quantum Mpemba effect. We recall that the spectrum of the Hamiltonian is listed in descending order, ${\varepsilon_1} > {\varepsilon_2}$. Consequently, the eigenvalues of the steady state are naturally arranged in ascending order, that is, $\Sigma = {U_1^{\dagger}}{\texttt{R}_1}{U_1} = \text{diag}({p_-},{p_+})$, with ${p_-} < {p_+}$, for all $T > 0$. In this case, we verify that the spectrum of the initial state must be sorted in descending order to maximize its distance from equilibrium evaluated using the Hilbert-Schmidt distance, trace distance, and quantum relative entropy. To do so, we consider the permutation matrix ${P_{\pi}} = |{0}\rangle\langle{1}| + |{1}\rangle\langle{0}|$ such that ${P_{\pi}}{D}{P_{\pi}^{\dagger}} = \text{diag}(1,0) = |0\rangle\langle{0}|$. Therefore, the transformed state becomes [see Eq.~\eqref{eq:0000009}]
\begin{equation}
\label{eq:0000031}
{\rho'}({t_0}) = {U_1}{P_{\pi}}{D}{P_{\pi}^{\dagger}}{U_1^{\dagger}} = |1\rangle\langle{1}| ~.
\end{equation}
It can be shown that the dressed initial state in Eq.~\eqref{eq:0000031} leads to the suppression of the slowest decaying mode, i.e., $\text{Tr}({\texttt{L}_2}{\rho'}({t_0})) = \langle{1}|{U_1}{\texttt{L}'_2}{U_1^{\dagger}}|{1}\rangle = 0$ and $\text{Tr}({\texttt{L}_3}{\rho'}({t_0})) = \langle{1}|{U_1}{\texttt{L}'_3}{U_1^{\dagger}}|{1}\rangle = 0$. This ensures that the relaxation process is accelerated.

To investigate the quantum Mpemba effect, we analy\-ti\-cally evaluate the instantaneous states $\rho(t) = {e^{t\mathcal{L}}}[{\rho}({t_0})]$ and ${\rho'}(t) = {e^{t\mathcal{L}}}[{\rho'}({t_0})]$. On the one hand, we find that $\rho(t) = (1/2)(\mathbb{I} + \vec{r}(t)\cdot\vec{\sigma})$, where the Bloch vector $\vec{r}(t) = ({r_x}(t),{r_y}(t),{r_z}(t))$ has the time-dependent components
\begin{align}
\label{eq:0000032}
{r_x}(t) &= - {e^{-\frac{\gamma{t}}{2}\text{coth}\left(\frac{\delta}{2{k_B}T}\right)}}\sin(\delta{t}) ~, \\
\label{eq:0000033}
{r_y}(t) &= - {e^{-\frac{\gamma{t}}{2}\text{coth}\left(\frac{\delta}{2{k_B}T}\right)}}\cos(\delta{t}) ~, \\
\label{eq:0000034}
{r_z}(t) &= \left(1 - {e^{-\gamma{t}\text{coth}\left(\frac{\delta}{2{k_B}T}\right)}}\right)\text{tanh}\left(\frac{\delta}{2{k_B}T}\right) ~.
\end{align}
Here $\vec{\sigma} = ({\sigma_x},{\sigma_y},{\sigma_z})$ is the vector of Pauli matrices. On the other hand, we obtain ${\rho'}(t) = {e^{t\mathcal{L}}}[{\rho'}({t_0})] = (1/2)(\mathbb{I} + {\vec{r}{\,'}}(t)\cdot\vec{\sigma})$, with the time-dependent Bloch vector ${\vec{r}{\,'}}(t) = ({r'_x}(t),{r'_y}(t),{r'_z}(t))$, where
\begin{align}
\label{eq:0000035}
{r'_x}(t) &= {r'_y}(t) = 0 ~,\\
\label{eq:0000036}
{r'_z}(t) &= \frac{2{e^{\frac{\delta}{{k_B}T}}}\left(1 - {e^{-\gamma{t}\,\text{coth}\left(\frac{\delta}{2{k_B}T}\right)}}\right)}{{e^{\frac{\delta}{{k_B}T}}} + 1} - 1 ~.
\end{align}
In turn, the steady state can be recast as ${\texttt{R}_1} = (1/2)(\mathbb{I} + {\vec{r}_{\text{ss}}}\cdot\vec{\sigma})$, with the time-independent Bloch vector ${\vec{r}_{\text{ss}}} = ({r_{\text{ss}}^x},{r_{\text{ss}}^y},{r_{\text{ss}}^z})$, where
\begin{equation}
\label{eq:0000037}
{r_{\text{ss}}^x} = {r_{\text{ss}}^y} = 0 ~,\quad {r_{\text{ss}}^z} = \text{tanh}\left(\frac{\delta}{2{k_B}T}\right) ~.
\end{equation}
Next we address the Hilbert-Schmidt distance, trace distance, and quantum relative entropy for these states. The analytical expressions for the TD are given by
\begin{align}
\label{eq:0000038}
&{\mathcal{D}_{\text{TD}}}(\rho(t),{\texttt{R}_1}) = \frac{1}{2}{e^{-\frac{{\gamma}{t}}{2}\text{coth}\left(\frac{\delta}{2{k_B}T}\right)}}\times \nonumber\\
&\sqrt{1 + {e^{-{\gamma}{t}\text{coth}\left(\frac{\delta}{2{k_B}T}\right)}}{\text{tanh}^2}\left(\frac{\delta}{2{k_B}T}\right)}
\end{align}
and 
\begin{equation}
\label{eq:0000039}
{\mathcal{D}_{\text{TD}}}({\rho'}(t),{\texttt{R}_1}) = \frac{{e^{-{\gamma}{t}\,\text{coth}\left(\frac{\delta}{2{k_B}T}\right)}}{\text{sech}^2}\left(\frac{\delta}{2{k_B}T}\right)}{2\left(1 - \text{tanh}\left(\frac{\delta}{2{k_B}T}\right)\right)} ~.
\end{equation}
We find that ${\mathcal{D}_{\text{HSD}}}(\rho(t),{\texttt{R}_1}) = \sqrt{2}\,{\mathcal{D}_{\text{TD}}}(\rho(t),{\texttt{R}_1})$, and likewise ${\mathcal{D}_{\text{HSD}}}({\rho'}(t),{\texttt{R}_1}) = \sqrt{2}\,{\mathcal{D}_{\text{TD}}}({\rho'}(t),{\texttt{R}_1})$. This relation is consistent with the known properties of the HSD and TD measures for single-qubit states, as discussed in Ref.~\cite{PhysRevA.100.022103}. The analytical expressions for the relative entropies $S(\rho(t),{\texttt{R}_1})$ and $S({\rho'}(t),{\texttt{R}_1})$ are too long to be reported here. We note that the QRE is evaluated with the help of the expression~\cite{arXiv:2509.20347}
\begin{align}
\label{eq:0000040}
&S(\varrho,{\texttt{R}_1}) = \|\vec{u}\,\|\, \text{ln}\left(\sqrt{\frac{1 + \|\vec{u}\,\|}{1 - \|\vec{u}\,\|}} \,\right) \nonumber\\
&+ \text{ln}\left(\sqrt{\frac{1 - {\|\vec{u}\,\|^2}}{1 - {\|{\vec{r}_{\text{ss}}}\|^2}}}\,\right) - \text{ln}\left(\sqrt{\frac{1 + \|{\vec{r}_{\text{ss}}}\|}{1 - \|{\vec{r}_{\text{ss}}}\|}}\right)\frac{(\vec{u}\cdot{\vec{r}_{\text{ss}}})}{\|{\vec{r}_{\text{ss}}}\|} ~,
\end{align}
which holds for any single-qubit state $\varrho = (1/2)(\mathbb{I} + \vec{u}\cdot\vec{\sigma})$, and $\|\bullet\|$ stands for the Euclidean norm. We recall that the Bloch vectors $\vec{r}(t)$, ${\vec{r}{\,'}}(t)$, and ${\vec{r}_{\text{ss}}}$ were described in Eqs.~\eqref{eq:0000032}--\eqref{eq:0000037}.

\begin{figure}[!t]
\begin{center}
\includegraphics[scale=0.575]{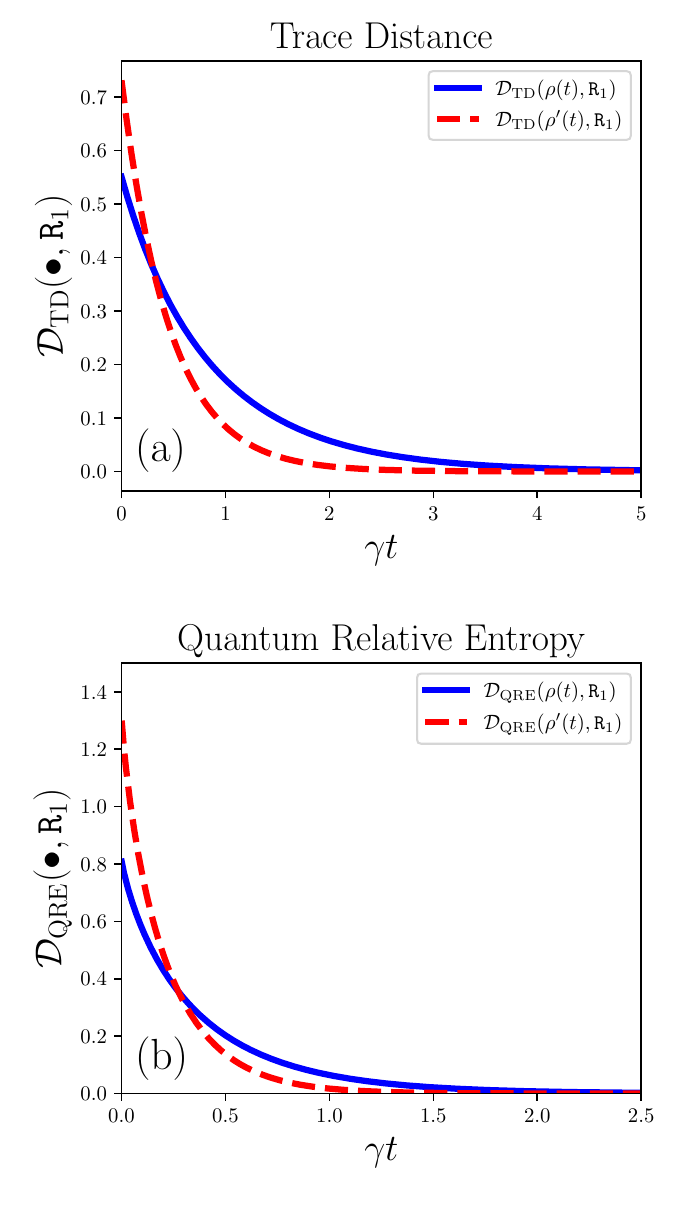}
\caption{(Color online) Quantum Mpemba effect in the dynamics of the two-level system described in Sec.~\ref{sec:00000000005A}. Here we set $\delta = {\varepsilon_1} - {\varepsilon_2} = 1$, $T = 1$, and ${k_B} = 1$. We investigate the relaxation of the instantaneous states ${\rho}(t) = {e^{t\mathcal{L}}}[{\rho}({t_0})]$ and ${\rho'}(t) = {e^{t\mathcal{L}}}[{\rho'}({t_0})]$, with ${\rho'}({t_0}) = U\rho({t_0}){U^{\dagger}}$, toward the steady state ${\texttt{R}_1}$ as a function of the dimensionless parameter $\gamma{t}$. We consider the following figures of merit: (a) the trace distance ${\mathcal{D}_{\text{TD}}}(\bullet,{\texttt{R}_1})$, and (b) the quantum relative entropy ${\mathcal{D}_{\text{QRE}}}(\bullet,{\texttt{R}_1})$. The blue solid lines correspond to ${\mathcal{D}_{\text{TD,QRE}}}(\rho(t),{\texttt{R}_1})$, while the red dashed lines represent ${\mathcal{D}_{\text{TD,QRE}}}({\rho'}(t),{\texttt{R}_1})$.}
\label{fig:FIG00002}
\end{center}
\end{figure}

In Fig.~\ref{fig:FIG00002}, we show the plots for the trace distance [see Fig.~\ref{fig:FIG00002}(a)] and quantum relative entropy [see Fig.~\ref{fig:FIG00002}(b)] related to the states $\rho(t)$, ${\rho'}(t)$, and $\texttt{R}_1$, as a function of the dimensionless parameter $\gamma{t}$. We set the parameters $\delta = {\varepsilon_1} - {\varepsilon_2} = 1$, $T = 1$, and ${k_B} = 1$. The plots clearly show that the relaxation to equilibrium is accelerated as a consequence of using the transformed state ${\rho'}({t_0})$. The plots show the occurrence of the quantum Mpemba effect, which is testified by the crossover between the relaxation curves for the selected distinguishability measures.


\begin{figure*}[!t]
\begin{center}
\includegraphics[scale=0.45]{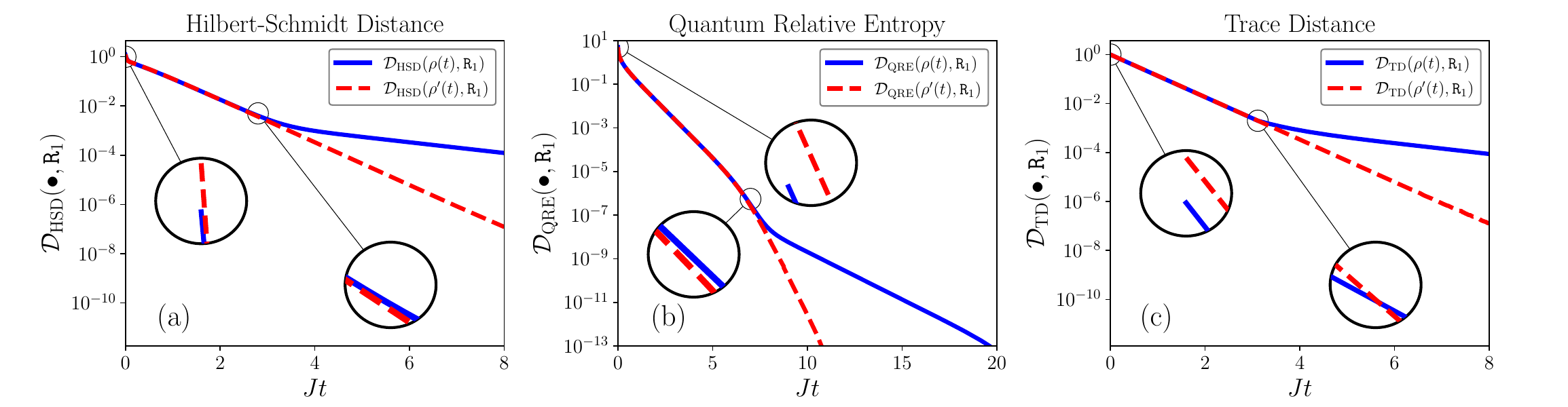}
\caption{(Color online) Quantum Mpemba effect in the dynamics of the transverse-field Ising model with open boundary conditions [see details in Sec.~\ref{sec:00000000005B}], where $N = 5$ spins, $h = J/2$, $\gamma = 1$, and $T = 0.1$. We investigate the relaxation of the instantaneous states ${\rho}(t) = {e^{t\mathcal{L}}}[{\rho}({t_0})]$ and ${\rho'}(t) = {e^{t\mathcal{L}}}[{\rho'}({t_0})]$, with ${\rho'}({t_0}) = U\rho({t_0}){U^{\dagger}}$, toward the steady state ${\texttt{R}_1}$ as a function of the dimensionless parameter $Jt$. We consider the following figures of merit: (a) the Hilbert-Schmidt distance ${\mathcal{D}_{\text{HSD}}}(\bullet,{\texttt{R}_1})$, (b) the quantum relative entropy ${\mathcal{D}_{\text{QRE}}}(\bullet,{\texttt{R}_1})$, and (c) the trace distance, ${\mathcal{D}_{\text{TD}}}(\bullet,{\texttt{R}_1})$. The blue solid lines correspond to ${\mathcal{D}_{\text{HSD,QRE,TD}}}(\rho(t),{\texttt{R}_1})$, while the red dashed lines represent ${\mathcal{D}_{\text{HSD,QRE,TD}}}({\rho'}(t),{\texttt{R}_1})$.}
\label{fig:FIG00003}
\end{center}
\end{figure*}

\subsection{Many-body quantum systems}
\label{sec:00000000005B}

In the following, we illustrate our findings for the QME by investigating the nonunitary dynamics of two paradigmatic spin models with open boundary conditions, namely, the transverse-field Ising (TFI) model
\begin{equation}
\label{eq:0000041}
{H_{\text{TFI}}} = - {J}\,{\sum_{j = 1}^{N - 1}}{\sigma_j^z}{\sigma_{j + 1}^z} + {h}\,{\sum_{j = 1}^N}{\sigma_j^x} ~,
\end{equation}
and the XXZ model
\begin{equation}
\label{eq:0000042}
{H_{\text{XXZ}}} = {\sum_{j = 1}^{N - 1}}[J({\sigma_j^x}{\sigma_{j + 1}^x} + {\sigma_j^y}{\sigma_{j + 1}^y}) + \Delta{\sigma_j^z}{\sigma_{j + 1}^z}] ~.
\end{equation}
Here ${\sigma_j^{x,y,z}}$ denotes the Pauli matrices acting on the $j$-th site of the chains, $N$ is the total number of spins, $J$ is the coupling between neighboring spins, $h$ corresponds to the strength of the transverse magnetic field, and $\Delta$ is the anisotropy parameter. These spin chains are weakly coupled to a bosonic reservoir at temperature $T$. The nonunitary dynamics of each system is governed by the Davies map discussed in Sec.~\ref{sec:00000000002}. 

For both models, the Hamiltonian spectrum is arranged in descending order, ${\varepsilon_1} > \ldots > {\varepsilon_d}$, whereas the corresponding steady state ${\texttt{R}_1} = {U_1}\Sigma{U_1^{\dagger}}$ has the spectrum listed in ascending order, with $\Sigma = \exp(-\varepsilon/{k_B}T)/Z$ and $Z = \text{Tr}(\exp(-\varepsilon/{k_B}T))$, where $\varepsilon = \text{diag}({\varepsilon_1},\ldots,{\varepsilon_d})$. The initial state is defined as $\rho({t_0}) = {U_1}|\phi\rangle\langle\phi|{U_1^{\dagger}}$, where $U_1$ is the matrix composed of the eigenvectors of the Hamiltonian for each specific spin model and $|\phi\rangle = (1/\sqrt{d}){\sum_{j = 1}^d}|j\rangle$, with $\{|j\rangle\}_{j = 1,\ldots,d}$ denoting the computational basis. Notably, the initial state differs between the two models, as $U_1$ depends on the specific Hamiltonian. Nevertheless, in both cases the state is pure, with its spectrum encoded by the same matrix $D = \text{diag}(0,0,\ldots,0,1)$. The matrix ${P_{\pi}} = {\sigma_x^{\otimes{N}}}$ is obtained by permuting the rows and columns of $D$ so that the spectrum of the probe state is arranged in descending order for both spin mo\-dels, i.e., ${P_{\pi}}D{P_{\pi}^{\dagger}} = \text{diag}(1,0\ldots,0,0)$. We have verified that this choice ensures that ${\mathcal{D}_{\text{HSD,QRE,TD}}}({\rho'}({t_0}),{\texttt{R}_1}) > {\mathcal{D}_{\text{HSD,QRE,TD}}}({\rho}({t_0}),{\texttt{R}_1})$ for all three distinguishability measures considered, and for both spin models.

\begin{figure*}[!t]
\begin{center}
\includegraphics[scale=0.45]{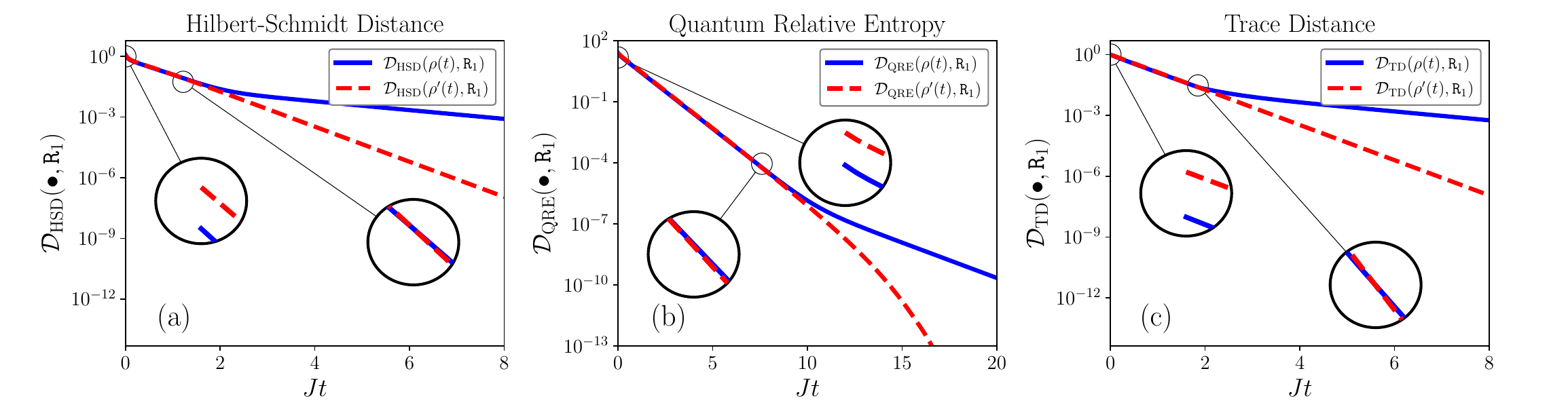}
\caption{(Color online) Quantum Mpemba effect in the dynamics of the XXZ model with open boundary conditions [see details in Sec.~\ref{sec:00000000005B}], where $N = 5$ spins, $\Delta = J/2$, $\gamma = 1$, and $T = 0.1$. We investigate the relaxation of the instantaneous states ${\rho}(t) = {e^{t\mathcal{L}}}[{\rho}({t_0})]$ and ${\rho'}(t) = {e^{t\mathcal{L}}}[{\rho'}({t_0})]$, with ${\rho'}({t_0}) = U\rho({t_0}){U^{\dagger}}$, toward the steady state ${\texttt{R}_1}$ as a function of the dimensionless parameter $Jt$. We consider the following figures of merit: (a) the Hilbert-Schmidt distance ${\mathcal{D}_{\text{HSD}}}(\bullet,{\texttt{R}_1})$, (b) the quantum relative entropy ${\mathcal{D}_{\text{QRE}}}(\bullet,{\texttt{R}_1})$, and (c) the trace distance, ${\mathcal{D}_{\text{TD}}}(\bullet,{\texttt{R}_1})$. The blue solid lines correspond to ${\mathcal{D}_{\text{HSD,QRE,TD}}}(\rho(t),{\texttt{R}_1})$, while the red dashed lines represent ${\mathcal{D}_{\text{HSD,QRE,TD}}}({\rho'}(t),{\texttt{R}_1})$.}
\label{fig:FIG00004}
\end{center}
\end{figure*}

Figures~\ref{fig:FIG00003} and~\ref{fig:FIG00004} present the numerical simulations for the TFI and XXZ models, respectively. We set $N = 5$, $h = J/2$, $\Delta = J/2$, $\gamma = 1$, $T = 0.1$, and ${k_B} = 1$. In these figures we show the plots for Hilbert-Schmidt distance [see Figs.~\ref{fig:FIG00003}(a) and~\ref{fig:FIG00004}(a)], quantum relative entropy [see Figs.~\ref{fig:FIG00003}(b) and~\ref{fig:FIG00004}(b)], and trace distance [see Figs.~\ref{fig:FIG00003}(c) and~\ref{fig:FIG00004}(c)], as a function of the dimensionless parameter $Jt$. The blue solid lines refer to ${\mathcal{D}_{\text{HSD},\text{QRE},\text{TD}}}(\rho(t),{\texttt{R}_1})$, with $\rho(t) = {e^{t\mathcal{L}}}[{\rho}({t_0})]$, while ${\mathcal{D}_{\text{HSD},\text{QRE},\text{TD}}}({\rho'}(t),{\texttt{R}_1})$ is represented by red dashed lines, with ${\rho'}(t) = {e^{t\mathcal{L}}}[{\rho'}({t_0})]$.

The insets in Figs.~\ref{fig:FIG00003} and~\ref{fig:FIG00004} show the crossovers that clearly indicate the occurrence of the genuine Mpemba effect for each spin model. For the TFI model, we find the crossover to occur at (i) $J{t_{\text{QME}}} \approx 1.47$ for the HSD [see Fig.~\ref{fig:FIG00003}(a)], (ii) $J{t_{\text{QME}}} \approx 6.16$ for the QRE [see Fig.~\ref{fig:FIG00003}(b)], and (iii) $J{t_{\text{QME}}} \approx 3.11$ for the TD [see Fig.~\ref{fig:FIG00003}(c)]. For the XXZ model, the corresponding times are (i) $J{t_{\text{QME}}} \approx 1.23$ for the HSD [see Fig.~\ref{fig:FIG00004}(a)], (ii) $J{t_{\text{QME}}} \approx 7.45$ for the QRE [see Fig.~\ref{fig:FIG00004}(b)], and (iii) $J{t_{\text{QME}}} \approx 1.85$ for the TD [see Fig.~\ref{fig:FIG00004}(c)]. The plots for the Hilbert-Schmidt distance and trace distance display closely similar behavior for both models, as illustrated in Figs.~\ref{fig:FIG00003}(a) and~\ref{fig:FIG00003}(c) for the TFI model, and in Figs.~\ref{fig:FIG00004}(a) and~\ref{fig:FIG00004}(c) for the XXZ model. The insets further indicate that the HSD [see Figs.~\ref{fig:FIG00003}(a) and~\ref{fig:FIG00004}(a)] exhibits a more pronounced decay at earlier times ($Jt \ll 1$) than the TD [Figs.~\ref{fig:FIG00003}(c) and~\ref{fig:FIG00004}(c)].

Notably, although the HSD, QRE and TD can exhibit distinct behaviors, thus leading to different relaxation dynamics, they are not independent, but rather obey well-established bounds. In particular, the TD and HSD satisfy the inequality $(1/\sqrt{2}){\mathcal{D}_{\text{HSD}}}(\rho,\varrho) \leq {\mathcal{D}_{\text{TD}}}(\rho,\varrho) \leq \sqrt{{\text{rank}(\rho)\text{rank}(\varrho)}/{[\text{rank}(\rho) + \text{rank}(\varrho)]}} \, {\mathcal{D}_{\text{HSD}}}(\rho,\varrho)$~\cite{PhysRevA.100.022103}, where $\text{rank}(\bullet)$ denotes the rank of the density matrix. Furthermore, the QRE and TD are related via the Pinsker inequality ${\mathcal{D}_{\text{QRE}}}(\rho,\varrho) \geq 2 {\mathcal{D}^2_{\text{TD}}}(\rho,\varrho)$~\cite{arXiv:2601.10395}. Combining these results yields $(1/\sqrt{2}){\mathcal{D}_{\text{HSD}}}(\rho,\varrho) \leq {\mathcal{D}_{\text{TD}}}(\rho,\varrho) \leq (1/\sqrt{2})\sqrt{{\mathcal{D}_{\text{QRE}}}(\rho,\varrho)}$. The results presented in Figs.~\ref{fig:FIG00003} and~\ref{fig:FIG00004} are consistent with this inequality. We emphasize that the variation in the critical times ${t_{\text{QME}}}$ across the three measures indicates that, for the spin models considered, the QRE is less sensitive than the HSD and TD in capturing the signatures of the genuine quantum Mpemba effect. Consequently, the crossover between relaxation curves occurs on longer timescales when quantified using quantum relative entropy. We have also verified that, beyond the choice of distinguishability measure, the value of ${t_{\text{QME}}}$ depends on both the system size and the initial state $\rho({t_0})$, which may further influence the saturation of these bounds.


\section{Accelerated relaxation and quantum coherences}
\label{sec:00000000006}

We proved that the exponential speedup of the dynamics of a given Davies map depends on the fact that the transformed state ${\rho'}({t_0}) = U\rho({t_0}){U^{\dagger}}$ is incoherent with respect to the eigenbasis ${\{|{\psi_l}\rangle\}_{l = 1,\ldots,d}}$ of the Hamiltonian $H$. This suggests that, for a probe state $\rho({t_0})$ that is already incoherent in this preferred basis, one can na\-tu\-rally expect faster relaxation to equilibrium. This property, however, can also be observed for a probe state that exhibit coherence in such a basis. Here we argue that there are initial states which spontaneously lead to exponentially fast relaxation to the equilibrium, but they need not be incoherent in the eigenbasis of the Hamiltonian of the open quantum system.

Let $\rho({t_0})$ be an initial state of an open quantum system described by a Davies map. This state may or may not be incoherent in the energy eigenbasis. We recall that ${\texttt{L}_2} = {U_1}{\texttt{L}'_2}{U_1^{\dagger}}$, where ${\texttt{L}'_2} = |{k_0}\rangle\langle{l_0}|$ is a non-Hermitian matrix that has a single nonzero element related to the pair $({k_0},{l_0}) \in \{1,2,\ldots,d\}$, with ${k_0} \neq {l_0}$. Then, the overlap between ${\texttt{L}_2}$ and $\rho({t_0})$ becomes
\begin{align}
\label{eq:0000043}
\text{Tr}({\texttt{L}_2}\rho({t_0})) &= \text{Tr}({\texttt{L}'_2}{U_1^{\dagger}}\rho({t_0}){U_1}) \nonumber\\
&= \langle{l_0}|{U_1^{\dagger}}\rho({t_0}){U_1}|{k_0}\rangle \nonumber\\
&= \langle{\psi_{l_0}}|\rho({t_0})|{\psi_{k_0}}\rangle ~,
\end{align}
where we used the fact that $|{\psi_s}\rangle = {U_1}|{s}\rangle$ is the $s$-th eigenstate of $H$.

It is worth noting that Eq.~\eqref{eq:0000043} allows for three possible scenarios. First, a vanishing overlap is obtained whenever $\rho({t_0})$ is an incoherent state in the energy eigenbasis. This means that the relaxation to equilibrium is expected to be accelerated. Second, for an initial state with nonzero coherences in this preferred basis, particularly with respect to the states $|{l_0}\rangle$ and $|{k_0}\rangle$, we obtain $\text{Tr}({\texttt{L}_2}\rho({t_0})) = \langle{\psi_{l_0}}|\rho({t_0})|{\psi_{k_0}}\rangle \neq 0$. In this case, faster relaxation can be achieved by using the dressed state ${\rho'}({t_0}) = U\rho({t_0}){U^{\dagger}}$ as an input to the nonunitary evolution. Finally, there may exist a probe state $\rho({t_0})$ that exhibit nonzero coherences for all eigenstates of the Hamiltonian, except for the vector pair $|{l_0}\rangle$ and $|{k_0}\rangle$, i.e., $\text{Tr}({\texttt{L}_2}\rho({t_0})) = \langle{\psi_{l_0}}|\rho({t_0})|{\psi_{k_0}}\rangle = 0$. To verify this, it is sufficient for matrix ${U_1^{\dagger}}\rho({t_0}){U_1}$ to have an off-diagonal entry equal to zero at the coordinate $({l_0},{k_0})$, while the only nonzero off-diagonal entry of matrix ${\texttt{L}'_2}$ is located at $({k_0},{l_0})$.

This class of incoherent states spontaneously suppresses the slowest decaying mode of the Liouvillian generator associated with the complex eigenvalues $\lambda_2$ and ${\lambda^*_2}$. Consequently, the state $\rho(t) = {e^{t\mathcal{L}}}[\rho({t_0})]$ relaxes rapidly toward equilibrium, with a shortened timescale $1/|\text{Re}({\lambda_3})|$, with $0 \leq |\text{Re}({\lambda_2})| \leq |\text{Re}({\lambda_3})| \leq \ldots \leq |\text{Re}({\lambda_{d^2}})|$. According to Ref.~\cite{PhysRevLett.127.060401}, such a speedup is sufficient to constitute the quantum Mpemba effect for this class of states. However, this does not characterizes a genuine quantum Mpemba effect as defined by Ref.~\cite{PhysRevLett.133.140404}. Notably, this effect can be engineered by the protocol described in Secs.~\ref{sec:00000000003} and~\ref{sec:00000000004}. Specifically, we consider two states $\rho(t) = {e^{t\mathcal{L}}}[\rho({t_0})]$ and ${\rho'}(t) = {e^{t\mathcal{L}}}[{\rho'}({t_0})] = {e^{t\mathcal{L}}}[{U}\rho({t_0}){U^{\dagger}}]$. The unitary $U$ is expected to suppress the contribution of the actual slowest decaying mode $\lambda_3$, provided it belongs to the subset of $d(d - 1)$ complex eigenvalues of the Liouvillian superoperator. Once this mode is eliminated, the system approaches equilibrium exponentially fast. The genuine quantum Mpemba effect occurs if, for some time $t > {t_0}$, one has $\mathcal{D}({\rho'}(t),{\texttt{R}_1}) < \mathcal{D}({\rho}(t),{\texttt{R}_1})$, provided $\mathcal{D}({\rho'}({t_0}),{\texttt{R}_1}) > \mathcal{D}({\rho}({t_0}),{\texttt{R}_1})$ at time ${t_0}$, for a given distinguishability measure.

To illustrate these findings, we consider a system of two qubits with the Hamiltonian $H = {U_1}\varepsilon{U_1^{\dagger}}$, with $\varepsilon = \text{diag}({\varepsilon_1},\ldots,{\varepsilon_4})$, where the energies are listed in descending order, ${\varepsilon_j} > {\varepsilon_{j + 1}}$ for $j = \{1,2,3\}$. Here ${U_1} = |{0,+}\rangle\langle{0,0}| + |{0,-}\rangle\langle{0,1}| + |{1,0}\rangle\langle{1,0}| + |{1,1}\rangle\langle{1,1}|$ is the unitary that brings the Hamiltonian of the system into its diagonal form, with $|{\pm}\rangle = (1/\sqrt{2})(|{0}\rangle \pm |{1}\rangle)$. Let ${\texttt{L}'_2} = |{0,1}\rangle\langle{0,0}|$ be the eigenoperator related to the slowest decaying mode of the Liouvillian. On the one hand, for the initial state $\rho({t_0}) = |{0,0}\rangle\langle{0,0}|$, we have that ${U_1^{\dagger}}\rho({t_0}){U_1} = |{0,+}\rangle\langle{0,+}|$, and we obtain the nonzero off-diagonal ele\-ment $\langle{0,0}|{U_1^{\dagger}}\rho({t_0}){U_1}|{0,1}\rangle = 1/2$. In this case, the dynamics of the quantum system can be accelerated exponentially by mapping ${\rho}({t_0})$ to the dressed state ${\rho'}({t_0})$, according to our protocol. On the other hand, when choosing the probe state $\rho({t_0}) = |{1,+}\rangle\langle{1,+}|$, we obtain ${U_1^{\dagger}}\rho({t_0}){U_1} = |{1,+}\rangle\langle{1,+}|$, which implies a zero-value off-diagonal element $\langle{0,0}|{U_1^{\dagger}}\rho({t_0}){U_1}|{0,1}\rangle = 0$. Hence, it follows that the dynamics of the quantum system is already exponentially accelerated, even though $\rho({t_0})$ is not incoherent with respect to the energy eigenbasis.


\section{Conclusion}
\label{sec:00000000007}

We have presented a framework to ge\-ne\-rate a genuine quantum Mpemba effect in the relaxation of open quantum systems. The protocol implements unitary operations built from permutation matrices that cause a rearrangement of the spectrum of a given initial state. On the one hand, the slower decay mode of the Liouville operator is suppressed, which triggers an exponentially faster relaxation to the equilibrium. On the other hand, such a protocol maps an input state to an incoherent state in the energy eigenbasis that is as far away as possible from the steady state of the dynamics. Once these conditions are met, we guarantee that a genuine Mpemba effect will occur. 

To cha\-rac\-terize the relaxation process, we monitor the distinguishability between instantaneous states and the steady state of the open quantum system. To do so, we consider the Hilbert-Schmidt distance, quantum relative entropy, and trace distance. We have shown that, for any initial state, there always exists a permutation matrix that maximizes the distance from the dressed state to equilibrium. In contrast to the results in Ref.~\cite{PhysRevLett.133.140404}, our protocol significantly reduces the computational cost of numerical simulations. We emphasize that our framework can reproduce the results of Ref.~\cite{PhysRevLett.127.060401}, with appropriate modifications and refinements to certain assumptions [see Appendix~\ref{sec:0000000000A}]. 

We focus on dissipative processes that exclude contributions from jump operators corresponding to zero-frequency transitions ($m = n$), which represent pure dephasing and leave populations unchanged. The interplay between relaxation dynamics and the quantum Mpemba effect in quantum master equations that account all Bohr frequencies of the system merits further investigation. We hope that our results will inspire work in this direction. Our results contribute to elucidate the mechanism underlying the quantum Mpemba effect and pave the way for insightful perspectives for the study of the dynamics of open quantum systems.


\begin{acknowledgments}
This work was supported by the Brazilian ministries MEC and MCTIC, and the Brazilian funding agencies CNPq, and Coordena\c{c}\~{a}o de Aperfei\c{c}oamento de Pessoal de N\'{i}vel Superior--Brasil (CAPES) (Finance Code 001). D. P. P. would like to acknowledge the Funda\c{c}\~{a}o de Amparo \`{a} Pesquisa e ao Desenvolvimento Cient\'{i}fico e Tecnol\'{o}gico do Maranh\~{a}o (FAPEMA) (Edital Acordo de Coopera\c{c}\~{a}o T\'{e}cnica - Bolsas Produtividade em Pesquisa Estaduais FAPEMA/CNPq, Grant No.~PQ-C-12651/25).
\end{acknowledgments}

\setcounter{equation}{0}
\setcounter{table}{0}
\setcounter{section}{0}
\numberwithin{equation}{section}
\makeatletter
\renewcommand{\thesection}{\Alph{section}} 
\renewcommand{\thesubsection}{\Alph{section}.\arabic{subsection}}
\def\@gobbleappendixname#1\csname thesubsection\endcsname{\Alph{section}.\arabic{subsection}}
\renewcommand{\theequation}{\Alph{section}\arabic{equation}}
\renewcommand{\thefigure}{\arabic{figure}}
\renewcommand{\bibnumfmt}[1]{[#1]}
\renewcommand{\citenumfont}[1]{#1}

\section*{Appendix}


\section{Exponentially accelerated relaxation for the real slowest decaying mode}
\label{sec:0000000000A}

Here we show that the results from Sec.~\ref{sec:00000000003} can be properly modified to retrieve those addressed in Ref.~\cite{PhysRevLett.127.060401}. We consider the dynamics of an open quantum system coupled to a Markovian environment described by the Markovian master equation $d\rho(t)/dt = \mathcal{L}[\rho(t)]$, where $\rho(t)$ is the instantaneous state of the system for all $t \geq 0$, while $\mathcal{L}[\bullet]$ is the Lindblad superoperator~\cite{Breuer_Petruccione}
\begin{equation}
\label{eq:appendixA0001}
\mathcal{L}[\bullet] = -i [H,\bullet] + {\sum_{j = 1}^N}\left({L_j}\bullet{L^{\dagger}_j} - \frac{1}{2}\left\{{L^{\dagger}_j}{L_j},\bullet\right\}\right) ~.
\end{equation}
Here $H$ is the Hamiltonian of the system, which is Hermitian, while the set $\{{L_j}\}_{j = 1,\ldots,N}$ of non-Hermitian jump operators characterizes the dissipative effects induced by the environment. This superoperator generates a completely positive dynamics, being trace-preserving, $\text{Tr}(\mathcal{L}[\bullet]) = 0$, and Hermitian, ${(\mathcal{L}[\bullet])^{\dagger}} = \mathcal{L}[{\bullet^{\dagger}}]$~\cite{Rivas_Huelga_book}.

Here we set the lowest decaying mode ${\texttt{L}_2}$ of the Liouvillian to be Hermitian, i.e., ${\texttt{L}^{\dagger}_2} = {\texttt{L}_2}$, with $\lambda_2$ a real eigenvalue. On the one hand, the Hermitian eigenmatrix admits the Schur decomposition ${\texttt{L}_2} = {U_1}{\texttt{L}'_2}{U_1^{\dagger}}$, where ${\texttt{L}'_2} = \text{diag}({\zeta_1},\ldots,{\zeta_d})$ is a diagonal matrix that takes into account the set of real eigenvalues of ${\texttt{L}_2}$, and ${U_1}$ is a unitary matrix formed by the respective eigenstates~\cite{artin2014algebra}. On the other hand, the pure initial state of the system is written as $\rho({t_0}) = \Lambda{D}{\Lambda^{\dagger}}$, where $D = \text{diag}(0,\ldots,1,\ldots,0)$ is a diagonal matrix whose elements are the eigenvalues of the probe state, while $\Lambda$ is a unitary matrix formed by the respective eigenstates.

Next we consider the probe state $\rho({t_0})$ mapped to ${\rho'}(t_0) = U\rho({t_0}){U^{\dagger}}$ by means of the unitary operator $U = {U_1}V(\theta){\Lambda^{\dagger}}$. Here we appropriately choose $V(\theta)$ as the unitary matrix
\begin{equation}
\label{eq:appendixA0002}
V(\theta) = {e^{i\theta {S_{\pi}}}}{P_{\pi}} = \left(\cos\theta \, {\mathbb{I}_d} + i\sin\theta{S_{\pi}}\right){P_{\pi}} ~,
\end{equation}
with $\theta \in [0,\pi/2]$, while ${S_{\pi}}$ and ${P_{\pi}}$ are real-value permutation matrices. We note that ${S_{\pi}^{\top}} = {S_{\pi}}$ is a symmetric matrix, while ${P_{\pi}^{\dagger}} = {P_{\pi}^{-1}}$ is a unitary permutation matrix. In this setting, the overlap between the eigenmatrix ${\texttt{L}_2}$ and the transformed state $U\rho({t_0}){U^{\dagger}}$ is written as
\begin{align}
\label{eq:appendixA0003}
\text{Tr}\left({\texttt{L}_2}{U}\rho({t_0}){U^{\dagger}}\right) &= {\cos^2}\theta \, \text{Tr}({\texttt{L}'_2}{P_{\pi}}{D}{P_{\pi}^{\dagger}}) \nonumber\\ 
& + {\sin^2}\theta \, \text{Tr}({S_{\pi}^{\dagger}}{\texttt{L}'_2}{S_{\pi}}{P_{\pi}}{D}{P_{\pi}^{\dagger}}) \nonumber\\
&- \frac{i}{2}\sin(2\theta)\text{Tr}\left({S_{\pi}}[{\texttt{L}'_2},{P_{\pi}}{D}{P_{\pi}^{\dagger}}]\right) ~.
\end{align}
In the following, we will discuss how the elimination of the slowest decay mode can be achieved.

Let ${P_{\pi}}$ be an arbitrary permutation matrix. We note that the diagonal matrix $D$ is always mapped onto another diagonal matrix ${P_{\pi}}{D}{P_{\pi}^{\dagger}}$ under the action of some permutation matrix ${P_{\pi}}$. The overall effect is the rearrangement of the entries along the diagonal of the original matrix. Therefore, since the diagonal matrices ${\texttt{L}'_2}$ and ${P_{\pi}}{D}{P_{\pi}^{\dagger}}$ commute with each other, i.e., ${\texttt{L}'_2}{P_{\pi}}{D}{P_{\pi}^{\dagger}} = {P_{\pi}}{D}{P_{\pi}^{\dagger}}{\texttt{L}'_2}$, we can readily conclude that
\begin{equation}
\label{eq:appendixA0004}
\text{Tr}({S_{\pi}}[{\texttt{L}'_2},{P_{\pi}}{D}{P_{\pi}^{\dagger}}]) = 0 ~,
\end{equation}
regardless of the permutation matrix chosen. The remaining coefficients $\text{Tr}({\texttt{L}'_2}{P_{\pi}}{D}{P_{\pi}^{\dagger}})$ and $\text{Tr}({S_{\pi}^{\dagger}}{\texttt{L}'_2}{S_{\pi}}{P_{\pi}}{D}{P_{\pi}^{\dagger}})$ on the right-hand side of Eq.~\eqref{eq:appendixA0003} are not necessarily equal to zero. In fact, they are related to the eigenvalues of the operator ${\texttt{L}_2}$ and, for a given critical value $\theta_c$, they both can combine and cancel each other out, thus ensuring the overlap in Eq.~\eqref{eq:appendixA0003} is zero. To outline this proof, we follow here a similar technical discussion to that presented in Ref.~\cite{PhysRevLett.127.060401}. We note that, for a suitable choice for the permutation matrix ${P_{\pi}}$, we have that the diagonal matrix $D$ is mapped onto ${P_{\pi}}{D}{P_{\pi}^{\dagger}} = \text{diag}(1,0,\ldots,0)$, which in turn implies that 
\begin{equation}
\label{eq:appendixA0005}
\text{Tr}({\texttt{L}'_2}{P_{\pi}}{D}{P_{\pi}^{\dagger}}) = {\zeta_1} ~. 
\end{equation}
To address the overlap $\text{Tr}({S_{\pi}^{\dagger}}{\texttt{L}'_2}{S_{\pi}}{P_{\pi}}{D}{P_{\pi}^{\dagger}})$, two points need to be highlighted. First, since the left eigenmatrix is Hermitian, it can be written as ${\texttt{L}_2} = {\sum_{k = 1}^d}\,{\zeta_k}|{\varphi_k}\rangle\langle{\varphi_k}|$, where ${\{|{\varphi_k}\rangle\}_{k = 1,\ldots,d}}$ is the set of eigenstates of ${\texttt{L}_2}$ that defines a basis for the Hilbert space, with $\langle{\varphi_k}|{\varphi_l}\rangle = {\delta_{kl}}$ and ${\sum_{k = 1}^d}\,|{\varphi_k}\rangle\langle{\varphi_k}| = {\mathbb{I}_d}$. Second, we recall that the eigenmatrices $\texttt{R}_1$ and $\texttt{L}_2$ are normalized to be biorthogonal to each other, that is, $\text{Tr}({\texttt{R}_1}{\texttt{L}_2}) = 0$. Hence, we obtain the constraint
\begin{equation}
\label{eq:appendixA0006}
{\sum_{k = 1}^d}\,{\zeta_k}\langle{\varphi_k}|{\texttt{R}_1}|{\varphi_k}\rangle = 0 ~.
\end{equation}
The right eigenmatrix ${\texttt{R}_1}$ is the steady state of the system, i.e., it is a positive-semidefinite matrix that satisfies $\langle{\varphi_k}|{\texttt{R}_1}|{\varphi_k}\rangle \geq 0$ for any eigenstate $|{\varphi_k}\rangle$ of the operator $\texttt{L}_2$. This means that, to satisfy Eq.~\eqref{eq:appendixA0006}, the set of eigenvalues cannot consist exclusively by positive real numbers, nor only of negative real numbers. In fact, there must be at least two nonzero eigenvalues with opposite sign, for example, the two eigenvalues $\zeta_1$ and $\zeta_n$ such that $\text{sgn}({\zeta_n}) = - \text{sgn}({\zeta_1})$. We then construct permutation matrices ${S_{\pi}}$ and ${P_{\pi}}$ to rearrange the ordering of the diagonal entries of matrices ${\texttt{L}'_2}$ and $D$, respectively, so that 
\begin{equation}
\label{eq:appendixA0007}
\text{Tr}({S_{\pi}^{\dagger}}{\texttt{L}'_2}{S_{\pi}}{P_{\pi}}{D}{P_{\pi}^{\dagger}}) = {\zeta_n} ~.
\end{equation}
Therefore, combining Eq.~\eqref{eq:appendixA0003} with Eqs.~\eqref{eq:appendixA0004},~\eqref{eq:appendixA0005}, and~\eqref{eq:appendixA0007}, we obtain the result
\begin{equation}
\label{eq:appendixA0008}
\text{Tr}\left({\texttt{L}_2}{U}\rho({t_0}){U^{\dagger}}\right) = {\zeta_1}{\cos^2}\theta + {\zeta_n}{\sin^2}\theta ~.
\end{equation}
The slowest decaying mode can be neglected if, for a given critical angle $\theta_c$, we have that $\text{Tr}\left({\texttt{L}_2}{U}\rho({t_0}){U^{\dagger}}\right) = 0$, which requires that
\begin{equation}
\label{eq:appendixA0009}
{\theta_c} = \arctan\left(\sqrt{\left|\frac{\zeta_1}{\zeta_n}\right|} \,\right) ~.
\end{equation}
We note that Eqs.~\eqref{eq:appendixA0008} and~\eqref{eq:appendixA0009} exactly recover the results obtained in Ref.~\cite{PhysRevLett.127.060401}.


\section{Proof of Eq.~$\eqref{eq:0000019}$}
\label{sec:0000000000B}

In this Appendix, we present the proof of Eq.~\eqref{eq:0000019}. The Hilbert-Schmidt distance probes the quantum Mpemba effect when the inequality 
\begin{equation}
\label{eq:0000004}
\text{Tr}({\rho'}({t_0}){\texttt{R}_1}) \leq \text{Tr}({\rho}({t_0}){\texttt{R}_1}) ~,
\end{equation} 
is satisfied, where ${\rho'}({t_0}) = U\rho({t_0}){U^{\dagger}}$ is a quantum state obtained from the initial state $\rho({t_0})$ of the system, $U$ is a unitary matrix, and ${\texttt{R}_1}$ is the steady state of the system. Let $U = {U_1}{P_{\pi}}{\Lambda^{\dagger}}$ be a unitary matrix, where ${P_{\pi}}$ is an arbitrary permutation matrix, with ${P_{\pi}^{\dagger}} = {P_{\pi}^{-1}}$. Here $U_1$ and $\Lambda$ are unitary matrices formed by the eigenstates of the Hamiltonian $H$ and the probe state $\rho({t_0})$, respectively. The steady state for the Davies map is given by the Gibbs state, ${\texttt{R}_1} = \exp(-\beta{H})/Z = {U_1}\Sigma{U_1^{\dagger}}$, with $\Sigma = \exp(-\varepsilon/{k_B}T)/Z$, where $\varepsilon$ is the diagonal matrix formed by the eigenvalues of $H$ and $Z = \text{Tr}(\exp(-\varepsilon/{k_B}T))$ is the partition function. In this case, we have that 
\begin{align}
\label{eq:appendixB0002}
\text{Tr}({\rho'}({t_0}){\texttt{R}_1}) &= {Z^{-1}}\text{Tr}({U_1^{\dagger}}{U_1}{P_{\pi}}{\Lambda^{\dagger}}\Lambda{D}{\Lambda^{\dagger}}\Lambda{P_{\pi}^{\dagger}}{U_1^{\dagger}}{U_1}\Sigma) \nonumber\\
&= {Z^{-1}}\text{Tr}({P_{\pi}}D{P_{\pi}^{\dagger}}\Sigma) ~, 
\end{align}
where we have used the cyclic property of the trace, while
\begin{align}
\label{eq:appendixB0003}
\text{Tr}({\rho}({t_0}){\texttt{R}_1}) &= {Z^{-1}}\text{Tr}({U_1^{\dagger}}\Lambda{D}{\Lambda^{\dagger}}{U_1}\Sigma) \nonumber\\
&= {Z^{-1}}\text{Tr}({U'}{D}{U'^{\dagger}}\Sigma) ~, 
\end{align}
where we define $U' := {U_1^{\dagger}}\Lambda$. Hence, by substituting Eqs.~\eqref{eq:appendixB0002} and~\eqref{eq:appendixB0003} into Eq.~\eqref{eq:0000004}, we arrive at the inequality
\begin{equation}
\label{eq:appendixB0004}
\text{Tr}({P_{\pi}}D{P_{\pi}^{\dagger}}\Sigma) \leq \text{Tr}({U'}{D}{U'^{\dagger}}\Sigma) ~.
\end{equation}
To be clear, the role of the permutation matrix ${P_{\pi}}$ is to rearrange the entries along the diagonal of the matrix $D$. In detail, given $D = \text{diag}({q_1},\ldots,{q_d})$, we have that ${P_{\pi}}{D}{P_{\pi}^{\dagger}} = \text{diag}({q_{\pi(1)}},\ldots,{q_{\pi(d)}})$, where $\pi(j) = k$ defines the permutation operation mapping a given number $j$ into $k$, with $j,k\in\{1,\ldots,d\}$.

Next we use that $\Sigma = {\sum_j}\,{\alpha_j}|{j}\rangle\langle{j}|$ and $D = {\sum_j}\,{q_j}|{j}\rangle\langle{j}|$, where ${\{|{j}\rangle\}_{j = 1,\ldots,d}}$ is the computational basis. In this case, the term $\text{Tr}({U'}{D}{U'^{\dagger}}\Sigma)$ in the right-hand side of Eq.~\eqref{eq:appendixB0004} becomes
\begin{equation}
\label{eq:0000011}
\text{Tr}({U'}{D}{U'^{\dagger}}\Sigma) = {\sum_j}\,{\alpha_j}{\sum_k}\,{q_k}{M_{jk}} ~,
\end{equation}
where we define
\begin{equation}
\label{eq:0000013}
{M_{jk}} := {|\langle{j}|{U'}|{k}\rangle|^2} ~.
\end{equation}
We recognize ${M_{jk}} = \langle{j}|M|{k}\rangle$ as the $(j,k)$-th element of bistochastic matrix, since it fulfills the property ${\sum_j}\,{M_{jk}} = {\sum_k}\,{M_{jk}} = 1$. Indeed, by performing the sum over index $j$, we have that
\begin{equation}
\label{eq:0000014}
{\sum_j}\,{M_{jk}} = {\sum_j}\,{|\langle{j}|{U'}|{k}\rangle|^2} = {\sum_j}\,\langle{k}|{U'^{\dagger}}|{j}\rangle\langle{j}|{U'}|{k}\rangle = 1 ~,
\end{equation}
and also the sum over index $k$,
\begin{equation}
\label{eq:appendixB0008}
{\sum_k}\,{M_{jk}} = {\sum_k}\,{|\langle{j}|{U'}|{k}\rangle|^2} = {\sum_k}\,\langle{j}|{U'}|{k}\rangle\langle{k}|{U'^{\dagger}}|{j}\rangle = 1 ~,
\end{equation}
where we have used that ${\sum_j}\,|{j}\rangle\langle{j}| = \mathbb{I}$. The same result holds for the sum over index $k$. In this case, according to the Birkhoff--von Neumann theorem, the matrix $M$ is recast in terms of a convex combination of permutation matrices~\cite{ANDO1989163}, i.e.,
\begin{equation}
\label{eq:appendixB0009}
M = {\sum_l}\,{\eta_l}{A_l} ~,
\end{equation}
where $A_l$ is a given permutation matrix, with $0 \leq {\eta_l} \leq 1$ and ${\sum_l}\,{\eta_l} = 1$. In this case, we obtain
\begin{align}
\label{eq:appendixB0010}
\text{Tr}({U'}{D}{U'^{\dagger}}\Sigma) &= {\sum_l}\,{\eta_l} \, {\sum_{j,k}}\,{\alpha_j}{q_k}\langle{j}|{A_l}|{k}\rangle \nonumber\\
&= {\sum_l}\,{\eta_l} \, {\sum_{j,k}}\,{\alpha_j}{q_k}{|\langle{j}|{A_l}|{k}\rangle|^2} ~,
\end{align}
where we used the fact that ${|\langle{j}|{A_l}|{k}\rangle|^2} = \langle{j}|{A_l}|{k}\rangle|$ since ${A_l}$ is a permutation matrix. It can be readily proved that
\begin{align}
\label{eq:appendixB0011}
{\sum_{j,k}}\,{\alpha_j}{q_k}{|\langle{j}|{A_l}|{k}\rangle|^2} &= {\sum_{j,k}}\,{\alpha_j}{q_k}\langle{j}|{A_l}|{k}\rangle\langle{k}|{A_l^{\dagger}}|{j}\rangle\nonumber\\
&= \text{Tr}({A_l}{D}{A_l^{\dagger}}\Sigma) ~.
\end{align}
Hence, by combining Eqs.~\eqref{eq:appendixB0010} and~\eqref{eq:appendixB0011}, we obtain the result
\begin{equation}
\label{eq:appendixB0012}
\text{Tr}({U'}{D}{U'^{\dagger}}\Sigma) = {\sum_l}\,{\eta_l} \text{Tr}({A_l}{D}{A_l^{\dagger}}\Sigma)  ~.
\end{equation}
Finally, by combining Eqs.~\eqref{eq:appendixB0004} and~\eqref{eq:appendixB0012}, we obtain
\begin{equation}
\label{eq:appendixB0013}
\text{Tr}({P_{\pi}}{D}{P_{\pi}^{\dagger}}\Sigma) \leq {\sum_l}\,{\eta_l}\text{Tr}({A_l}{D}{A_l^{\dagger}}\Sigma) ~.
\end{equation}

The right-hand side of Eq.~\eqref{eq:appendixB0013} defines a convex sum ${\sum_l}\,{\eta_l}{c_l}$ of the non-negative elements ${c_l} := \text{Tr}({A_l}{D}{A_l^{\dagger}}\Sigma)$. We note that ${c_{\text{min}}} \leq {\sum_l}\,{\eta_l}{c_l} \leq {c_{\text{max}}}$, where ${c_{\text{min}}} = \text{min}\{{c_1},\ldots,{c_d}\}$ and ${c_{\text{max}}} = \text{max}\{{c_1},\ldots,{c_d}\}$ for all $0 \leq {\eta_l} \leq 1$, with ${\sum_l}\,{\eta_l} = 1$. This means that there exist two permutation matrices ${P_{\pi,\text{max}}}$ and ${P_{\pi,\text{min}}}$ that satisfy the upper bounds $\text{Tr}({P_{\pi,\text{max}}}{D}{P_{\pi,\text{max}}^{\dagger}}\Sigma) \leq {c_{\text{max}}}$ and $\text{Tr}({P_{\pi,\text{min}}}{D}{P_{\pi,\text{min}}^{\dagger}}\Sigma) \leq {c_{\text{min}}}$, respectively. The inequality in Eq.~\eqref{eq:appendixB0013} is satisfied in both scenarios, but we realize that the latter case provides a tighter upper bound than the former. In particular, there must exist an optimal permutation matrix ${P_{\pi,\text{opt}}}$ such that $\text{Tr}({P_{\pi,\text{opt}}}{D}{P_{\pi,\text{opt}}^{\dagger}}\Sigma) \leq {c_{\text{min}}}$. To see this point, we note that 
\begin{equation}
\label{eq:appendixB0014}
\text{Tr}({P_{\pi,\text{opt}}}{D}{P_{\pi,\text{opt}}^{\dagger}}\Sigma) = {\sum_{j = 1}^d}\,{\alpha_j}{q_{\pi(j)}} ~, 
\end{equation}
where $\pi(\bullet)$ maps index $j$ to $\pi(j)$, with $j,\pi(j) \in \{1,\ldots,d\}$. Therefore, to ensure that ${\sum_{j = 1}^d}\,{\alpha_j}{q_{\pi(j)}} \leq {c_{\text{min}}}$, it suffices to choose a permutation matrix such that ${P_{\pi,\text{opt}}}{D}{P^{\dagger}_{\pi,\text{opt}}} = \text{diag}({q_{\pi(1)}},\ldots,{q_{\pi(d)}})$, with ${q_{\pi(1)}} \geq \ldots \geq {q_{\pi(d)}}$ listed in descending order provided ${\alpha_1} \leq \ldots \leq {\alpha_n}$ are listed in ascending order, or vice versa. Indeed, it is known from the so-called rearrangement inequality that ${\sum_{j = 1}^d}\,{\alpha_j}{q_{\pi(j)}}$ takes its minimum value if ${\alpha_1} \leq \ldots \leq {\alpha_d}$ and ${q_{\pi(1)}} \geq \ldots \geq {q_{\pi(d)}}$~\cite{hardy1952inequalities}. This proves the validity of the inequality in Eq.~\eqref{eq:appendixB0013}, which also guarantees that Eq.~\eqref{eq:0000004} is valid.



%
 
\end{document}